\documentclass[12pt]{article}
\pdfoutput=1
\usepackage{latexsym,amsmath,amssymb, graphicx, subfigure}
\usepackage[linktocpage]{hyperref}
\renewcommand{\thefootnote}{\fnsymbol{footnote}}

\usepackage{mathrsfs }
\usepackage{amsmath}
\usepackage{amsfonts}
\usepackage{tikz}

\numberwithin{equation}{section}

\def\doubleset#1#2{\bgroup%
\def\doit#1#2{%
\setbox\dblsetbox=\hbox{$\cstyle #1$}%
\raise#2\ht\dblsetbox\copy\dblsetbox%
\hskip-\wd\dblsetbox%
\raise-#2\ht\dblsetbox\box\dblsetbox}%
\mathchoice%
{\def\cstyle{\displaystyle}\doit#1#2}%
{\def\cstyle{\textstyle}\doit#1#2}%
{\def\cstyle{\scriptstyle}\doit#1#2}%
{\def\cstyle{\scriptscriptstyle}\doit#1#2}\egroup}
\def\underarrow#1{\vbox{\ialign{##\crcr$\hfil\displaystyle
 {#1}\hfil$\crcr\noalign{\kern1pt\nointerlineskip}$\longrightarrow$\crcr}}}

\newbox\dblsetbox

\newlength{\extraspace}
\newlength{\extraspaces}
\newcommand{\be}{\begin{equation}
\addtolength{\abovedisplayskip}{\extraspaces}
\addtolength{\belowdisplayskip}{\extraspaces}
\addtolength{\abovedisplayshortskip}{\extraspace}
\addtolength{\belowdisplayshortskip}{\extraspace}}
\newcommand{\ee}{\end{equation}}

\newcommand{\ba}{\begin{eqnarray}
\addtolength{\abovedisplayskip}{\extraspaces}
\addtolength{\belowdisplayskip}{\extraspaces}
\addtolength{\abovedisplayshortskip}{\extraspace}
\addtolength{\belowdisplayshortskip}{\extraspace}}
\newcommand{\ea}{\end{eqnarray}}

\newcommand{\bd}{\begin{displaymath}
\addtolength{\abovedisplayskip}{\extraspaces}
\addtolength{\belowdisplayskip}{\extraspaces}
\addtolength{\abovedisplayshortskip}{\extraspace}
\addtolength{\belowdisplayshortskip}{\extraspace}}
\newcommand{\ed}{\end{displaymath}}

\newcounter{saveeqn}

\newcommand{\newsection}[1]{
\vspace{12mm} \pagebreak[3] \addtocounter{section}{1}
\setcounter{equation}{0} \setcounter{subsection}{0}
\noindent{\bf \thesection. #1} \nopagebreak
\medskip
\nopagebreak
\addcontentsline{toc}{section}{\thesection. #1}}

\newcommand{\newsubsection}[1]{
\vspace{0.8cm} \pagebreak[3] \addtocounter{subsection}{1}
\setcounter{subsubsection}{0}
\noindent{ \it \thesubsection. #1} \nopagebreak \vspace{2mm}
\nopagebreak
\addcontentsline{toc}{subsection}{\thesubsection. #1}}

\begin{document}
\addtolength{\baselineskip}{1.5mm}

\thispagestyle{empty}

\vbox{} \vspace{-0.0cm}

\begin{center}
\centerline{\Large{\bf An M-Theoretic Derivation of a 5d and 6d AGT}}
\medskip
\centerline{\Large{\bf Correspondence, and Relativistic and Elliptized}}
\medskip
\centerline{\Large{\bf Integrable Systems}}

\vspace{2.0cm}

{\bf{Meng-Chwan~Tan}}
\\[0mm]
{\it Department of Physics,
National University of Singapore}\\[0 mm]
mctan@nus.edu.sg
\end{center}

\vspace{2.0 cm}

\centerline{\bf Abstract}\smallskip

We generalize our analysis in [arXiv:1301.1977], and show that a 5d and 6d AGT correspondence for $SU(N)$ --  which essentially relates the relevant 5d and 6d Nekrasov instanton partition functions to the integrable representations of a $q$-deformed and elliptic affine ${\cal W}_N$-algebra -- can be derived, purely physically, from the principle that the spacetime BPS spectra of \emph{string-dual} M-theory compactifications ought to be equivalent. Via an appropriate defect, we also derive a ``fully-ramified'' version of the 5d and 6d AGT correspondence where integrable representations of a quantum and elliptic affine $SU(N)$-algebra at the critical level appear  on the 2d side, and argue that the relevant ``fully-ramified'' 5d and 6d Nekrasov instanton partition functions are simultaneous eigenfunctions of commuting operators which define relativistic and elliptized integrable systems. As an offshoot, we also obtain various mathematically novel and interesting relations involving the double loop algebra of $SU(N)$, elliptic Macdonald operators, equivariant elliptic genus of instanton moduli space, and more.

\newpage

\renewcommand{\thefootnote}{\arabic{footnote}}
\setcounter{footnote}{0}

\tableofcontents

\newsection{Introduction, Summary and Acknowledgements}

In 2009, Alday-Gaiotto-Tachikawa verified in~\cite{AGT} that the Nekrasov instanton partition function of a 4d ${\cal N} = 2$ conformal $SU(2)$ quiver theory is equivalent to a conformal block of a 2d CFT with ${\cal W}_2$-algebra symmetry that is Liouville theory. This celebrated 4d-2d correspondence, better known since as the AGT correspondence, was soon proposed to also hold for a 4d ${\cal N} = 2$  asymptotically-free $SU(2)$ theory~\cite{irr}, and a 4d ${\cal N} = 2$ conformal $SU(N)$ quiver theory where the corresponding 2d CFT is an $A_{N-1}$ conformal Toda field theory which has ${\cal W}_N$-algebra symmetry~\cite{AGT-matter}.

The basis for the AGT correspondence for $SU(N)$ -- as first pointed out in~\cite{Alday-Tachikawa} -- is a conjectured relation between the equivariant cohomology of the moduli space of $SU(N)$-instantons and the integrable representations of an affine ${\cal W}_N$-algebra. This conjectured relation was proved mathematically in~\cite{Vasserot, Maulik-Okounkov}, and physically derived in~\cite{4d AGT} via the principle that the spacetime BPS spectra  of \emph{string-dual} M-theory compactifications ought to be equivalent.   

``Ramified''  generalizations of the AGT correspondence for pure $SU(N)$ involving arbitrary surface operators were later proposed in~\cite{AGT-ram, Kanno-Tachikawa}, where they were also physically derived in~\cite{4d AGT} via the principle that the spacetime BPS spectra  of \emph{string-dual} M-theory compactifications ought to be equivalent. For a full surface operator,  the correspondence was also proved purely mathematically in~\cite{J-function}.

The AGT correspondence also implies certain relations between the Nekrasov instanton partition function and 2d integrable systems. An example  would be the conjecture in~\cite{Alday-Tachikawa}, which asserts that the ``fully-ramified'' Nekrasov instanton partition function should be related to Hitchin's integrable system. This was again physically derived using M-theory in~\cite{4d AGT}.

Our main aim is to furnish in a pedagogical manner, a fundamental M-theoretic derivation of a 5d and 6d analog of the above 4d-2d relations, by generalizing our analysis in~\cite{4d AGT}.  Let us now give a brief plan and summary of the paper.

\bigskip\noindent{\it A Brief Plan and Summary of the Paper}

 In $\S$2, we will first review the M-theoretic derivation of a 4d pure AGT correspondence in~\cite{4d AGT}. Then, we will generalize the analysis in \emph{loc.~cit.} to furnish an M-theoretic derivation of a 5d AGT correspondence for pure $SU(N)$ in the topological string limit. In particular, we find that the corresponding 5d Nekrasov instanton partition function can be expressed as the inner product of a coherent state in an integrable module over a universal central extension of the double loop algebra of $SU(N)$. Our result therefore sheds further light on an earlier field-theoretic result by Nekrasov-Okounkov in~\cite{NO} regarding the aforementioned 5d Nekrasov instanton partition function.

 In $\S$3, we will first review the M-theoretic derivation of a 4d AGT correspondence for a conformal linear quiver theory in~\cite{4d AGT}. Then, we will generalize the analysis in \emph{loc.~cit.} to furnish an M-theoretic derivation of a 5d AGT correspondence for $SU(N)$ with $N_f = 2N$ fundamental matter.  In starting with the $N=1$ case,  we will first make contact with what is known as a Ding-Iohara algebra, whence for the $N > 1$ case, we find that the corresponding 5d Nekrasov instanton partition function can be expressed as a four-point correlation function on a sphere of vertex operators of an integrable module over a $q$-deformed affine ${\cal W}_N$-algebra. Our result therefore serves as a purely physical M-theoretic proof of a mathematical conjecture by Awata-Feigin-Hoshino-Kanai-Shiraishi-Yanagida in~\cite{Awata}, of a 5d analog of the AGT correspondence for $SU(N)$ with $N_f = 2N$ fundamental matter. Last but not least, we will proceed in the same manner to furnish an M-theoretic derivation of a 5d AGT correspondence for pure $SU(N)$, where we find that the corresponding 5d Nekrasov instanton partition function -- which is also given by an equivariant index of the Dirac operator on the moduli space of $SU(N)$-instantons --  can be expressed as the inner product of a coherent state in an integrable module over a $q$-deformed affine ${\cal W}_N$-algebra -- which is also given by a Whittaker function on the  $q$-deformed affine ${\cal W}_N$-algebra. Our result therefore serves as an $SU(N)$ generalization of a 5d analog of the AGT correspondence for pure $SU(2)$ first proposed and checked by Awata-Yamada in~\cite{Awata-AGT}. Furthermore, as an offshoot of our physical analysis in the topological string limit, we also have the novel representation-theoretic result that a Whittaker vector in a level $N$ module over a (certain specialization of the) Ding-Iohara algebra is also a Whittaker vector in a module over a universal central extension of the double loop algebra of $SU(N)$.  
 
 In $\S$4, we will first review the M-theoretic derivation of a 4d ``fully-ramified'' pure AGT correspondence in~\cite{4d AGT}.  Then, we will generalize the analysis in \emph{loc.~cit.} to furnish an M-theoretic derivation of a 5d ``fully-ramified'' AGT correspondence for (i) pure $SU(N)$ and (ii) $SU(N)$ with single adjoint matter, in the Nekrasov-Shatashvilli (NS) limit. In particular, we find that the corresponding 5d Nekrasov instanton partition function can be expressed as (i) the inner product of a coherent state in an integrable module over a quantum affine $SU(N)$-algebra at the critical  level and (ii) a one-point correlation function on a torus of a vertex operator of an integrable module over a  quantum affine $SU(N)$-algebra at the critical  level. Along the way, we will also argue that the corresponding 5d Nekrasov instanton partition function ought to be a simultaneous eigenfunction of commuting operators which define a (i) relativistic periodic Toda integrable system and (ii) relativistic elliptic Calogero-Moser system. As an offshoot, we also have the interesting result that the simultaneous eigenfunctions of an elliptic generalization of the celebrated Macdonald operators can be understood as one-point correlation functions of a 2d QFT on a torus whose underlying symmetry is generated by a quantum affine $SU(N)$-algebra at the critical level.
 
 In $\S$5, we will generalize our analysis in $\S$3 to furnish  an M-theoretic derivation of a 6d  AGT correspondence for the non-anomalous case of an $SU(N)$ theory with $N_f = 2N$ fundamental matter. In starting with the $N=1$ case, we will first make contact with what is known as an elliptic Ding-Iohara algebra, whence for the $N > 1$ case, we find that the corresponding 6d Nekrasov instanton partition function -- which is also given by an equivariant elliptic genus on the moduli space of $SU(N)$-instantons -- can be expressed as a two-point correlation function on a torus of vertex operators of an integrable module over an elliptic affine ${\cal W}_N$-algebra.
 
 In $\S$6, we will generalize our analysis in $\S$4 to furnish an M-theoretic derivation of a 6d ``fully-ramified'' AGT correspondence for $SU(N)$ with $N_f = 2N$ fundamental matter in the NS limit.  In particular, we find that the corresponding 6d Nekrasov instanton partition function can be expressed as  a two-point correlation function on a torus of vertex operators of an integrable module over an elliptic affine $SU(N)$-algebra at the critical  level.  Along the way, we will also argue that the corresponding 6d Nekrasov instanton partition function ought to be a simultaneous eigenfunction of commuting operators which define an elliptized integrable system.

\bigskip\noindent{\it Acknowledgements} 

I would like to thank H.~Awata, H.~Konno, Y.~Saito, P.~Sulkowski and K.~Takemura for very helpful exchanges.  This work is supported in part by the NUS Startup Grant.

\vspace{-0.3cm}

\newsection{An M-Theoretic Derivation of a 5d Pure AGT Correspondence in the Topological String Limit}

\vspace{-0.3cm}

\newsubsection{An M-Theoretic Derivation of a 4d Pure AGT Correspondence: A Review}

We shall now review how an expected equivalence of the 6d spacetime BPS spectra of physically dual M-theory compactifications would allow us to derive a 4d AGT correspondence for pure $SU(N)$~\cite{4d AGT}. 

To this end, recall from \cite[(5.9) and (5.11)]{4d AGT} that we have the following \emph{physically dual} M-theory compactifications
\be
\underbrace{\mathbb R^4\vert_{\epsilon_1, \epsilon_2}  \times \Sigma_{t}}_{\textrm{$N$ M5-branes}}\times \mathbb R^{5}\vert_{\epsilon_3; \,  x_{6,7}}  \quad \Longleftrightarrow  \quad   {\mathbb R^{5}}\vert_{\epsilon_3; \, x_{4,5}} \times \underbrace{{\cal C}  \times TN_N^{R\to 0}\vert_{\epsilon_3; \, x_{6,7}}}_{\textrm{$1$ M5-branes}}.
\label{AGT-M-duality-AB}
\ee
Here, we have a common half-BPS boundary condition at the tips of the interval $\mathbb I_t \subset \Sigma_{t} = {\bf S}^1 \times \mathbb I_t$ (that is realized by  a pair of M9-branes which span all directions normal to them); the radius of ${\bf S}^1$ is $\beta$; $\mathbb I_t \ll \beta$; $\cal C$ is \emph{a priori} the same as $\Sigma_{t} $; and the $\epsilon_i$'s are parameters of the Omega-deformation along the indicated planes,\footnote{Thess parameters are real in our formulation, and for later convenience, we shall consider them to have the opposite sign.\label{parameters}} where the directions spanned by the M5-branes are:
\be
\begin{array}{l|c|c|cccc|cccc|c}
&0&1&2&3&4&5&6&7&8&9&10 \\
\hline
\hbox{$N$ M5's} & - & - & -& -& - & - &&&& & \\
\hbox{1 M5} & - & - & & &  &  & - &- &-& - & \\
\end{array} \label{table-AGT}
\ee
with $x_0$ and $x_1$ being the coordinates on $\mathbb I_t$ and ${\bf S}^1$, respectively. (Note that if $z = x_2 + i x_3$ and $w = x_4 + i x_5$, $\mathbb R^4$ can be viewed as a complex surface $\mathbb C^2$ whose coordinates are $(z,w)$. Likewise, if $u = x_6 + i x_7$ and $v = x_8 + i x_9$, $TN_N^{R\to 0}$ can be viewed as a complex surface whose singularity at the origin would be modeled by $\mathbb C^2 / \mathbb Z_N$, where $(u,v)$ are the coordinates on $\mathbb C^2$.)

\bigskip\noindent{\it The Spectrum of Spacetime BPS States on the LHS of (\ref{AGT-M-duality-AB})}

Let us first discuss the spectrum of spacetime BPS states along $\mathbb R^{5}\vert_{\epsilon_3; \,  x_{6,7}} \times \mathbb I_t$ on the LHS of (\ref{AGT-M-duality-AB}). In the absence of Omega-deformation whence $\epsilon_i = 0$, according to \cite[$\S$5.1]{4d AGT},  the spacetime BPS states would be captured by the topological sector of the ${\cal N} = (4,4)$ sigma-model on  $\Sigma_{t}$ with target the moduli space ${\cal M}_{SU(N)}$ of $SU(N)$-instantons on $\mathbb R^4$. However, in the presence of Omega-deformation, as explained in \cite[$\S$5.1]{4d AGT}, as one traverses a closed loop in $\Sigma_{t}$, there would be a  $\bf g$-automorphism of ${\cal M}_{SU(N)}$, where ${\bf g} \in U(1) \times U(1) \times T$, and $T \subset SU(N)$ is the Cartan subgroup. Consequently, the spacetime BPS states of interest would, in the presence of Omega-deformation, be captured by the topological sector of a non-dynamically ${\bf g}$-gauged version of the aforementioned sigma-model.\footnote{The relation between $\bf g$-automorphisms of the sigma-model target space and a non-dynamical $\bf g$-gauging of its worldsheet theory, is explained in~\cite[$\S$2.4 and $\S$5]{tsm}. For self-containment, let us review the idea here. Consider a sigma-model with worldsheet $\Sigma_{t}$, target space ${\cal M}_{SU(N)}$, and bosonic scalar fields $\Phi$. In the usual case where there is no action on ${\cal M}_{SU(N)}$ as we traverse a closed loop in $\Sigma_{t}$, one would consider in the sigma-model path-integral, the space of maps $\Phi : \Sigma_{t} \to {\cal M}_{SU(N)}$, which can be viewed as the space of trivial sections of a trivial bundle $X = {\cal M}_{SU(N)} \times \Sigma_{t}$. If however there is a  $\bf g$-automorphism of ${\cal M}_{SU(N)}$ as we traverse a closed loop in $\Sigma_{t}$, $X$ would have to be a nontrivial bundle given by $ {\cal M}_{SU(N)} \hookrightarrow X \to \Sigma_{t}$; then, $\Phi(z, \bar z)$ will not represent a map $\Sigma_{t} \to  {\cal M}_{SU(N)}$, but rather, it will be a nontrivial section of $X$. Thus, since $\Phi$ is no longer a function but a nontrivial section of a nontrivial bundle, its ordinary derivatives must be replaced by covariant derivatives. As the nontrivial structure group of $X$ is now $\bf g$, replacing ordinary derivatives by covariant derivatives would mean introducing on $\Sigma_{t}$ gauge fields $A^a$, which, locally, can be regarded as $({\rm Lie} \, {\bf g})$-valued one-forms with the usual gauge transformation law ${A^a}' = g^{-1} A^a g + g^{-1} dg$, where $g \in {\bf g}$. This is equivalent to gauging the sigma-model non-dynamically by $\bf g$. \label{gauging worldsheet}} Hence, according to~\cite{mine-equivariant} and our arguments in \cite[$\S$3.1]{4d AGT} which led us to \cite[eqn.~(3.5)]{4d AGT}, we can express the Hilbert space ${\cal H}^{\Omega}_{\rm BPS}$ of spacetime BPS states on the LHS of (\ref{AGT-M-duality-AB}) as
\be
{\cal H}^\Omega_{\rm BPS} = \bigoplus_{m} {\cal H}^\Omega_{{\rm BPS}, m}  =  \bigoplus_{m} ~{\rm IH}^\ast_{U(1)^2 \times T} \, {\cal U}({\cal M}_{SU(N), m}), 
\label{BPS-AGT-AB}
\ee
where ${\rm IH}^\ast_{U(1)^2 \times T} \, {\cal U}({\cal M}_{SU(N), m})$ is the $U(1)^2 \times T$-equivariant intersection cohomology of the Uhlenbeck compactification ${\cal U}({\cal M}_{SU(N), m})$ of the (singular) moduli space ${\cal M}_{SU(N), m}$ of $SU(N)$-instantons on $\mathbb R^4$ with instanton number $m$. 

Notably, as explained in \cite[$\S$5.1]{4d AGT}, the partition function of these spacetime BPS states would be given by the following 5d (brane) worldvolume expression 
\be
Z_{\rm BPS} (\epsilon_1, \epsilon_2, \vec a, \beta) =  \sum_m {\rm Tr}_{{\cal H}_m} \, {\rm exp} \, \beta (\epsilon_1 J_1 +  \epsilon_2 J_2 + {\vec a \cdot \vec T}),
\label{5d BPS}
\ee 
where $\vec T = (T_1 \dots, T_{{\rm rank} \, SU(N)})$ are the generators of the Cartan subgroup of $SU(N)$;  $\vec a = (a_1, \dots, a_{{\rm rank} \, SU(N)})$ are the corresponding purely imaginary Coulomb moduli of the $SU(N)$ gauge theory on $\mathbb R^4 \vert_{\epsilon_1, \epsilon_2}$; $J_{1, 2}$ are the rotation generators of the $x_2$-$x_3$ and $x_4$-$x_5$ planes, respectively, corrected with an appropriate amount of the $SU(2)_R$-symmetry to commute with the two surviving worldvolume supercharges; and ${\cal H}_m$ is the space of holomorphic functions on the moduli space ${\cal M}_{SU(N), m}$ of $SU(N)$-instantons on $\mathbb R^4$ with instanton number $m$.

\bigskip\noindent{\it The Spectrum of Spacetime BPS States on the RHS of (\ref{AGT-M-duality-AB})}

Let us next discuss the corresponding spectrum of spacetime BPS states along $\mathbb R^{5}\vert_{\epsilon_3; \, x_{4,5}} \times \mathbb I_t$ on the RHS of (\ref{AGT-M-duality-AB}). As explained in \cite[$\S$5.2]{4d AGT}, the spacetime BPS states would be furnished by the I-brane theory in the following type IIA configuration:
\be
\textrm{IIA}: \quad \underbrace{ {\mathbb R}^5\vert_{\epsilon_3; x_{4,5}} \times {\cal C} \times {\mathbb R}^3\vert_{\epsilon_3; x_{6,7}}}_{\textrm{I-brane on ${\cal C} = N \textrm{D6} \cap 1\textrm{D4}$}}.
\label{equivalent IIA system 1 - AGT}
\ee
Here, we have  a stack of $N$ coincident D6-branes whose worldvolume is given by ${\mathbb R}^5\vert_{\epsilon_3; x_{4,5}} \times {\cal C}$, and a single D4-brane whose worldvolume is given by ${\cal C} \times \mathbb R^3\vert_{\epsilon_3; x_{6,7}}$. 

Let us for a moment turn off Omega-deformation in (\ref{equivalent IIA system 1 - AGT}), i.e., let $\epsilon_3 = \epsilon_1 + \epsilon_2 = 0$. Then, as explained in \cite[$\S$5.2]{4d AGT},  the spacetime BPS states would be furnished by $N$ complex chiral fermions on ${\cal C} $ which effectively realize a  chiral WZW model at level 1 on ${\cal C} $, ${\rm WZW}^{{\rm level} \, 1}_{{{\frak {su}}(N)}_{{\rm aff}}}$, where ${{\frak {su}}(N)}_{{\rm aff}}$ is the affine $SU(N)$-algebra. 

Now turn Omega-deformation back on. As indicated in (\ref{equivalent IIA system 1 - AGT}), as one traverses around a closed loop in ${\cal C}$, the $x_4$-$x_5$ plane in ${\mathbb R^{4}}\vert_{\epsilon_3; \, x_{4,5}} \subset {\mathbb R^{5}}\vert_{\epsilon_3; \, x_{4,5}}$ would be rotated by an angle of $\epsilon_3$ together with an $SU(2)_R$-symmetry rotation of the supersymmetric $SU(N)$ gauge theory along $ {\mathbb R^{4}}\vert_{\epsilon_3; \, x_{4,5}}$. As such, Omega-deformation in this instance would effect a ${\bf g}'$-automorphism of ${\cal M}_{SU(N), l}$ as we traverse around a closed loop in ${\cal C}$, where ${\cal M}_{SU(N), l}$ is the moduli space of $SU(N)$-instantons on $\mathbb R^4$ with instanton number $l$; ${\bf g}' =  {\rm exp} \, \beta (\epsilon_3 J_3 + {{\vec a} \cdot {\vec T}'})$; $J_3$ is the rotation generator of the $x_4$-$x_5$ plane corrected with an appropriate amount of $SU(2)_R$-symmetry to commute with the D6-brane worldvolume supercharges; ${\vec T}' = (T'_1 \dots, T'_{{\rm rank} \, SU(N)})$ are the generators of the Cartan subgroup $T' \subset SU(N)$;  and ${\vec a} = (a_1, \dots, a_{{\rm rank} \, SU(N)})$ are the corresponding purely imaginary Coulomb moduli of the $SU(N)$ gauge theory on $\mathbb R^4 \vert_{\epsilon_3; \, x_{4,5}}$. In fact, since ${\cal M}_{SU(N), l}$ is also the space of self-dual connections of an $SU(N)$-bundle on $\mathbb R^4$, and since these self-dual connections correspond to differential one-forms valued in the Lie algebra $\frak {su}(N)$, Omega-deformation also means that there is a ${\bf g}'$-automorphism of the space of elements of $\frak {su}(N)$ and thus $SU(N)$,  as we traverse a closed loop in ${\cal C}$.

Note at this point that ${\rm WZW}^{{\rm level} \, 1}_{{{\frak {su}}(N)}_{{\rm aff}}}$ can be regarded as a (chiral half of a) $SU(N)$ WZW model at level 1 on ${\cal C}$. Since a $\cal G$ WZW model on $\Sigma$ is a bosonic sigma-model on $\Sigma$ with target the $\cal G$-manifold, according to the last paragraph,  it would mean that Omega-deformation would effect a ${\bf g}'$-automorphism  of the target space of ${\rm WZW}^{{\rm level} \, 1}_{{{\frak {su}}(N)}_{{\rm aff}}}$ as we traverse a closed loop in ${\cal C}$, where ${\bf g}' \in U(1) \times T'$.  In turn, according to footnote~\ref{gauging worldsheet}, it would mean that in the presence of Omega-deformation, we would have to non-dynamically gauge ${\rm WZW}^{{\rm level} \, 1}_{{{\frak {su}}(N)}_{{\rm aff}}}$ by $ U(1) \times T'$. 

That being said, notice also from (\ref{equivalent IIA system 1 - AGT}) that as one traverses around a closed loop in ${\cal C}$, the $x_6$-$x_7$ plane in ${\mathbb R^{3}}\vert_{\epsilon_3; \, x_{6,7}}$ would be rotated by an angle of $\epsilon_3$ together with an $R$-symmetry rotation of the supersymmetric $U(1)$ gauge theory living on the single D4-brane, i.e., Omega-deformation is also being turned on along the D4-brane. As explained in \cite[$\S$5.2]{4d AGT}, this means that we would in fact have to non-dynamically gauge ${\rm WZW}^{{\rm level} \, 1}_{{{\frak {su}}(N)}_{{\rm aff}}}$ not by $ U(1) \times T'$ but by ${\cal T} \subset T'$. 

At any rate, because $SU(N)/T' \simeq SL(N, \mathbb C) / B_+$, where $B_+$ is a Borel subgroup, it would mean that $SU(N)/{\cal T} \simeq (SL(N, \mathbb C) / B_+) \times (T' / \cal T)$. Also, $T' / \cal T$ is never bigger than the Cartan subgroup $C \subset B_+ = C \times N_+$, where $N_+$ is the subgroup of strictly upper triangular matrices which are nilpotent and traceless whose Lie algebra is $\frak n_+$. Altogether, this means that our gauged WZW model which corresponds to the coset model $SU(N)/{\cal T}$, can also be studied as an $S$-gauged  $SL(N, \mathbb C)$ WZW model which corresponds to the coset model  $SL(N, \mathbb C) / S$, where $N_+ \subseteq S \subset B_+$. As physically consistent $\cal H$-gauged $\cal G$ WZW models are such that $\cal H$ is necessarily a connected subgroup of $\cal G$, it will mean that $S = N_+$. Therefore, what we ought to ultimately consider is an $N_+$-gauged $SL(N, \mathbb C)$ WZW model. 

Before we proceed any further, let us make a slight deviation to highlight an important point regarding the effective geometry of ${\cal C}$. As the simple roots of $N_+$ form a subset of the simple roots of $SL(N,\mathbb C)$, the level of the affine $N_+$-algebra ought to be the equal to the level of the affine $SL(N,\mathbb C)$-algebra~\cite{Ketov} which is 1.  However, it is clear from our discussion hitherto that the affine $N_+$-algebra, in particular its level, will depend nontrivially on the Omega-deformation parameters which may or may not take integral values; in other words, its level will \emph{not} be equal to 1. A resolution to this conundrum is as follows. A deviation of the level of the affine $N_+$-algebra from 1 would translate into a corresponding deviation of its central charge; since a central charge arises due to an introduction of a macroscopic scale in the 2d system which results from a curvature along ${\cal C}$~\cite{CFT text}, it would mean that Omega-deformation ought to deform  the \emph{a priori} flat ${\cal C}  = \Sigma_{t}$ into a curved Riemann surface with the same topology -- that is, a Riemann sphere with two punctures -- such that the anomalous deviation in the central charge and therefore level, can be consistently ``absorbed'' in the process. Thus, we effectively have $ {\cal C} =  {\bf S}^2 / \{ 0, \infty \}$, so $\cal C$ can be viewed as an ${\bf S}^1$ fibration of $\mathbb I_t$ whose fiber has zero radius at the two end points $z = 0$ and $z = \infty$, where `$z$' is a holomorphic coordinate on $\cal C$.

Coming back to our main discussion, it is clear that in the schematic notation of \cite[$\S$3.1]{4d AGT}, our $N_+$-gauged $SL(N, \mathbb C)$ WZW model can be expressed as the partially gauged chiral CFT
\be
\frak {sl}(N)_{\rm aff, 1} / {\frak{n}_+}_{{\rm aff}, p}
\label{AGT-AB-chiral CFT}
\ee
on $\cal C$, where the level $p$ would, according to our discussions thus far, necessarily depend on the Omega-deformation parameters $\epsilon'_1 = \beta \epsilon_1$ and $\epsilon'_2 = \beta \epsilon_2$. ($p$, being a purely real number, should not depend on the purely imaginary parameter $\vec a' = \beta \vec a$).

In sum, the sought-after spacetime BPS states ought to be given by the states of the partially gauged chiral CFT in (\ref{AGT-AB-chiral CFT}), and via \cite[$\S$B]{4d AGT} and~\cite[eqn.~(6.67)]{review}, we find that this chiral CFT realizes an affine ${\cal W}_N$-algebra obtained from $\frak {sl}(N)_{\rm aff}$ via a quantum Drinfeld-Sokolov reduction. In other words,  the states of the chiral CFT would be furnished by a Verma module  $\widehat{{\cal W}}_N$ over the affine ${\cal W}_N$-algebra, and the Hilbert space ${\cal H}^{\Omega'}_{\rm BPS}$ of spacetime BPS states on the RHS of (\ref{AGT-M-duality-AB}) can be expressed as
\be
{\cal H}^{\Omega'}_{\rm BPS}  = \widehat{{\cal W}}_N.
\label{AGT-AB-H=W}
\ee

\bigskip\noindent{\it An AGT Correspondence for Pure $SU(N)$}

Clearly, the physical duality of the compactifications in (\ref{AGT-M-duality-AB}) will mean that ${\cal H}^\Omega_{\rm BPS}$ in (\ref{BPS-AGT-AB}) is equivalent to ${\cal H}^{\Omega'}_{\rm BPS}$ in (\ref{AGT-AB-H=W}), i.e.,
\be
\bigoplus_{m} ~{\rm IH}^\ast_{U(1)^2 \times T} \, {\cal U}({\cal M}_{SU(N), m}) = \widehat{{\cal W}}_N.
 \label{AGT-duality-A}
\ee
Moreover, as explained in \cite[$\S$5.2]{4d AGT}, the central charge and level of the affine ${\cal W}_N$-algebra are given by 
\be
\label{c-A-epsilon}
c_{A, \epsilon_{1,2} }  = (N-1)  + (N^3 - N) {(\epsilon_1 + \epsilon_2)^2 \over {\epsilon_1\epsilon_2}} \qquad {\rm and} \qquad  k'  = - N -  {\epsilon_2 / \epsilon_1},
\ee
respectively.

In the limit that $\beta \to 0$, it is well-known~\cite{abcd} that
 \be
 \label{abcd BPS relation}
 Z_{\rm BPS} (\epsilon_1, \epsilon_2, \vec a, \beta) =  \sum_m Z_{{\rm BPS}, m} (\epsilon_1, \epsilon_2, \vec a, \beta)
 \ee 
 of (\ref{5d BPS}) behaves as $Z_{{\rm BPS}, m} (\epsilon_1, \epsilon_2, \vec a, \beta \to 0)  \sim \beta^{-2mN} \, Z^{\rm 4d}_{{\rm BPS}, m} (\epsilon_1, \epsilon_2, \vec a)$, whence the 4d Nekrasov instanton partition function $Z_{\rm inst}  (\Lambda, \epsilon_1, \epsilon_2, \vec a)= \sum_m \Lambda^{2mN} \, Z^{\rm 4d}_{{\rm BPS}, m} (\epsilon_1, \epsilon_2, \vec a) $ can be written as 
\be
Z_{\rm inst}(\Lambda, \epsilon_1, \epsilon_2, \vec a) =  \sum_m \Lambda^{2mN} \, Z'_{{\rm BPS}, m} (\epsilon_1, \epsilon_2, \vec a, \beta \to 0), 
\label{Z_inst-5d}
\ee
where $Z'_{{\rm BPS}, m} = l_m \beta^{2mN} Z_{{\rm BPS}, m}$; $l_m$ is some constant; and $\Lambda$ can be interpreted as the inverse of the observed scale of the $\mathbb R^4\vert_{\epsilon_1, \epsilon_2}$ space on the LHS of (\ref{AGT-M-duality-AB}).

 Note that equivariant localization~\cite{Atiyah-Bott} implies that ${\rm IH}^\ast_{U(1)^2 \times T} \, {\cal U}({\cal M}_{SU(N), m})$ must be endowed with an orthogonal basis $\{| {{\vec p}_m} \rangle \}$, where the ${{\vec p}_m}$'s denote the fixed points of a $U(1)^2 \times T$-action on ${\cal U}({\cal M}_{SU(N), m})$.  Thus, since $Z'_{{\rm BPS}, m}$ is a weighted count of the states in ${\cal H}^\Omega_{{\rm BPS}, m} = {\rm IH}^\ast_{U(1)^2 \times T} \, {\cal U}({\cal M}_{SU(N), m})$, it would mean that one can write
\be
Z'_{{\rm BPS}, m} (\epsilon_1, \epsilon_2, \vec a, \beta \to 0) =  \sum_{{{\vec p}_m}} \,  l^2_{{{\vec p}_m}} (\epsilon_1, \epsilon_2, \vec a)\langle {{\vec p}_m} | {{\vec p}_m} \rangle,
\label{rhs of Z_inst-5d}
\ee
where $l_{{{\vec p}_m}} (\epsilon_1, \epsilon_2, \vec a) \in \mathbb R$, and the dependence on $\epsilon_1$, $\epsilon_2$ and $\vec a$ arises because the energy level of each state -- given by the eigenvalue $m$ of the $L_0$ operator which generates translation along ${\bf S}^1 \subset \Sigma_{t}$ in (\ref{AGT-M-duality-AB}) whence there is an Omega-deformation twist of the theory along the orthogonal spaces indicated therein -- ought to depend on these Omega-deformation parameters. 

Notice that (\ref{rhs of Z_inst-5d}) also means that
\be
Z'_{{\rm BPS}, m} (\epsilon_1, \epsilon_2, \vec a, \beta \to 0) =   \langle \Psi_m | \Psi_m \rangle,
\label{Z'}
\ee
where
\be
|\Psi_m \rangle = \bigoplus_{{{\vec p}_m}} \,  l_{{{\vec p}_m}}    | {{\vec p}_m} \rangle.
\ee 
Here, the state $| \Psi_m \rangle \in {\rm IH}^\ast_{U(1)^2 \times T} \, {\cal U}({\cal M}_{SU(N), m})$, and $ \langle \cdot | \cdot \rangle$ is a Poincar\'e pairing in the sense of~\cite[$\S$2.6]{J-function}.

Now consider the state
\be
\label{coherent-A}
| \Psi \rangle = \bigoplus_m  \Lambda^{m N} | \Psi_m \rangle.
\ee
By substituting (\ref{Z'}) in the RHS of (\ref{Z_inst-5d}), and by noting that $\langle \Psi_m | \Psi_n \rangle = \delta_{mn}$, one can immediately see that
\be
Z_{\rm inst} (\Lambda, \epsilon_1, \epsilon_2, \vec a) = \langle \Psi |  \Psi \rangle,
\label{Psi | Psi}
\ee
where $| \Psi \rangle \in \bigoplus_m {\rm IH}^\ast_{U(1)^2 \times T} \, {\cal U}({\cal M}_{SU(N), m})$.  In turn, the duality relation (\ref{AGT-duality-A}) would mean that 
\be
| \Psi \rangle = | \rm {coh} \rangle
\label{psi = delta}
\ee
whence  
\be
\label{q | q}
Z_{\rm inst} (\Lambda, \epsilon_1, \epsilon_2, \vec a) = \langle \rm {coh}   | \rm {coh} \rangle,
\ee
where $| {\rm {coh}} \rangle \in \widehat{{\cal W}}_N$. Since the RHS of (\ref{q | q}) is defined at $\beta \to 0$ (see the RHS of (\ref{Z_inst-5d})), and since we have in $\cal C$ a \emph{common} boundary condition at $z = 0$ and $z=\infty$,  $| \rm {coh} \rangle$ and $\langle \rm{coh} |$ ought to be a state and its dual associated with the puncture at $z = 0$ and $z=\infty$, respectively (as $z = 0, \infty$ are the points in $\cal C$ where the ${\bf S}^1$ fiber   has zero radius). Furthermore, as the RHS of (\ref{coherent-A}) is a sum over states of all possible energy levels, it would mean from (\ref{psi = delta}) that $| \rm {coh} \rangle$ is actually a \emph{coherent state}. This is depicted in fig.~1.  

\begin{figure}
  \centering
    \includegraphics[width=0.3\textwidth]{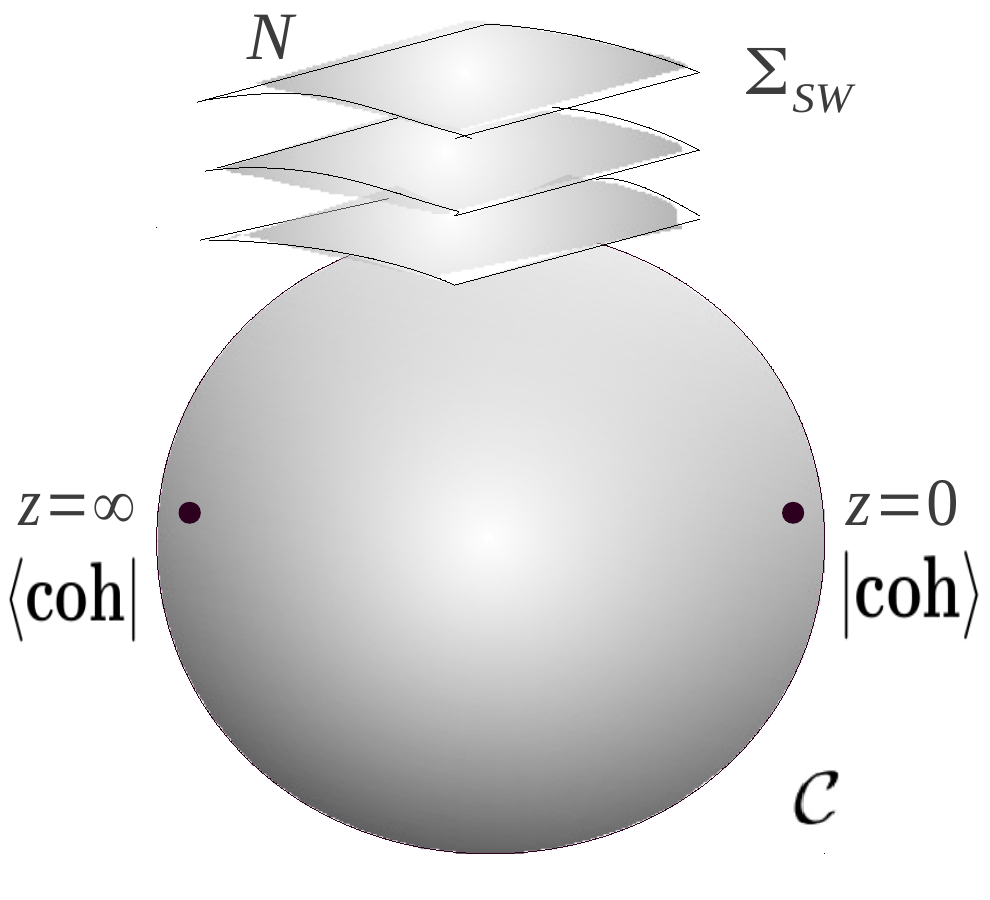}
  \caption{$\cal C$ and its $N$-fold cover $\Sigma_{SW}$ with the states $\langle \rm {coh} |$ and $| \rm {coh} \rangle$ at $z=0$ and $\infty$}
\end{figure}

At any rate, since we have $N$ D6-branes and $1$ D4-brane wrapping $\cal C$ (see (\ref{equivalent IIA system 1 - AGT})), we effectively have an $N \times 1 = N$-fold cover $\Sigma_{SW}$ of $\cal C$. This is also depicted in fig.~1. Incidentally, $\Sigma_{SW}$ is also the Seiberg-Witten curve which underlies $Z_{\rm inst} (\Lambda, \epsilon_1, \epsilon_2, \vec a)$! Moreover, it is by now well-established (see~\cite{Lerche} and references therein) that $\Sigma_{SW}$ can be described in terms of the algebraic relation 
\be
\label{4d SW curve}
\Sigma_{SW}: \lambda^N + \phi_2(z) \lambda^{N-2} + \dots + \phi_N(z) = 0, 
\ee
where $\lambda = y  dz /z$ (for some complex variable $y$) is a section of $T^\ast \cal C$; the $\phi_s(z)$'s are $(s,0)$-holomorphic differentials on $\cal C$ given by
 \be
 \label{phi(z)}
\phi_j (z)= u_j \left({dz \over z}\right)^j \quad {\rm and} \quad \phi_N(z) = \left( z + u_N + {\Lambda^N \over z}\right)\left({dz \over z}\right)^N,
\ee
where $j = 2, 3, \dots, N-1$; while for weights $\lambda_1, \dots, \lambda_N$ of the $N$-dimensional representation of $SU(N)$, and for $s=2, 3, \dots, N$,
$u_s = (-1)^{s+1} \sum_{k_1 \neq \dots \neq k_s} e_{\lambda_{k_1}} e_{\lambda_{k_2}} \dots e_{\lambda_{k_s}} (\vec a) \quad {\rm and} \quad e_{\lambda_{r}}  = \vec a \cdot \lambda_r$.  This is consistent with our results that  we have, on $\cal C$, the following $(s_i, 0)$-holomorphic differentials 
\be
\label{W's}
W^{(s_i)}(z) = \left(\sum_{l \in \mathbb Z} {W^{(s_i)}_l \over z^l} \right) \left( dz \over z \right)^{s_i}, \quad {\rm where} \quad s_i = e_i + 1 = 2, 3, \dots, N,
\ee
whence we can naturally identify, up to some constant factor,  $\phi_s (z)$ with  $W^{(s)}(z)$. (In fact, a $U(1)$ $R$-symmetry of the 4d theory along $\mathbb R^4\vert_{\epsilon_1, \epsilon_2}$ on the LHS of (\ref{AGT-M-duality-AB}) which underlies $Z_{\rm inst} (\Lambda, \epsilon_1, \epsilon_2, \vec a)$ and $\Sigma_{SW}$, can be identified with the rotational symmetry of ${\bf S}^1$; the duality relation (\ref{AGT-M-duality-AB}) then means that the corresponding $U(1)$ $R$-charge of the $\phi_s(z)$  operators that define  $\Sigma_{SW}$, ought to match, up to a constant, the conformal dimension of the $W^{(s)}(z)$ operators on $\cal C$, which is indeed the case.) 

In arriving at the relations (\ref{AGT-duality-A}), (\ref{c-A-epsilon}),  (\ref{psi = delta}) and (\ref{q | q}), and the identification between the $W^{(s_i)}$'s and the $\phi_j$'s,  we have just derived an AGT correspondence for pure $SU(N)$.

\newsubsection{An M-Theoretic Derivation of a 5d Pure AGT Correspondence in the Topological String Limit }

Let us now consider the topological string limit $\epsilon_1 + \epsilon_2 = \epsilon_3 =  0$ in our derivation of the 4d pure AGT correspondence in the last subsection. In this limit, Omega-deformation on the RHS of (\ref{AGT-M-duality-AB})  effectively vanishes. According to our discussions in the last subsection, (i) $\cal C$ in the I-brane configuration (\ref{equivalent IIA system 1 - AGT})  would become a flat finite cylinder $\Sigma_{t} = {\bf S}^1 \times \mathbb I_t$ again; (ii) the partially gauged chiral CFT behind (\ref{AGT-AB-chiral CFT}) would be \emph{ungauged}. This means that instead of (\ref{q | q}), we now have
\be
\label{q | q - hbar}
Z_{\rm inst} (\Lambda, \hbar,  \vec a)  =  \langle {u}_{\hbar}  | \Lambda^{2 N {\cal L}_0} | {u}_{\hbar} \rangle.
\ee
Here, $\Lambda$ is the energy scale; $\hbar = \epsilon_1 = - \epsilon_2$; $\vec a$ is the Coulomb moduli of the underlying 4d pure $SU(N)$ theory; $| u_{\hbar} \rangle \in {\widehat {{\frak {su}}(N)}_{{\rm aff}, 1}}$, where ${\widehat {{\frak {su}}(N)}_{{\rm aff}, 1}}$ is an integrable highest weight module over an  affine Lie algebra ${ {{\frak {su}}(N)}_{{\rm aff}, 1}}$ of level 1; $| u_{\hbar} \rangle$ is a \emph{coherent state} generated from the primary state $| \Delta_\hbar \rangle$ of conformal dimension $\Delta_\hbar \sim {\vec a}^2 / \hbar^2$;  and ${\cal L}_0$ is the generator of time translations along $\Sigma_{t}$ which propagates the state $| {u}_{\hbar} \rangle$ at one end by a distance $\sim 1 / g^2 \sim {\rm ln}  \, \Lambda^{2 N}$ to the other end whence it is annihilated by the state $\langle {u}_{\hbar} |$,  where $g$ is the underlying gauge coupling.\footnote{As the 4d gauge theory along $\mathbb R^4\vert_{\epsilon_1, \epsilon_2}$ in the original compactification in (\ref{AGT-M-duality-AB}) is, in this case, asymptotically-free, the observed scale of the eleven-dimensional spacetime $ \mathbb R^4\vert_{\epsilon_1, \epsilon_2} \times \Sigma_{t} \times \mathbb R^5$ ought to be inversely proportional to $g^2$; in particular, this means that the length of $\Sigma_{t}$ ought to be proportional to $1/g^2$.}

\bigskip\noindent{\it The 4d Nekrasov Instanton Partition Function and Complex Chiral Fermions}

Note that if  $\{e_l, h_l, f_l\}_{l = 0, \dots, N-1}$ are the Chevalley generators of ${ {{\frak {su}}(N)}_{{\rm aff}, 1}}$, where the $e_l$'s and $f_l$'s correspond to lowering and raising operators, respectively, we can write
\be
\label{uh-1}
| u_{\hbar} \rangle = {\rm {exp}} \left( \alpha \sum_{l=1}^N f_{l-1} \right) | v_0 \rangle,
\ee
where $\alpha$ is some normalization constant, and $| v_0 \rangle$ is the state corresponding to the highest weight vector $v_0$.

As mentioned in the last subsection, we have $N$ complex chiral fermions 
\be
\psi^{a}(z) = \sum_{r \in \mathbb Z + {1 \over 2}} \psi^a_r \, z^{-r} \, \left({dz \over z}\right)^{1 \over 2}, \quad {\tilde \psi}^{a}(z) = \sum_{r \in \mathbb Z + {1 \over 2}} {\tilde \psi}^a_r \, z^{-r} \, \left({dz \over z}\right)^{1 \over 2}, \quad a= 1, \dots N,
\ee
which realize $ {{\frak {su}}(N)}_{{\rm aff}, 1}$ on $\Sigma_{t}$. Consequently, by choosing $\alpha = N / \hbar$, we can also express (\ref{uh-1}) as (c.f.~\cite{NO})
\be
\label{uh-level 1}
| u_{\hbar} \rangle = e^{{\cal J}_{-1} \over \hbar} | p \rangle,
\ee
whence from (\ref{q | q - hbar}), the 4d Nekrasov instanton partition function would be given by
\be
\label{q | q - hbar - chiral fermions}
Z_{\rm inst} (\Lambda, \hbar, \vec a)  =  \langle p  | e^{{\cal J}_{1} \over \hbar} \, {\Lambda}^{2 N {\cal L}_0} \, e^{{\cal J}_{-1} \over \hbar} | p \rangle,
\ee 
where $| p \rangle$ is a vacuum state in a standard fermionic Fock space ${\cal H}_{[p]}$ whose energy level is $p^2 /2 \sim \vec a^2 / \hbar^2$, and ${\cal J}_{\mp n}$ are creation and annihilation operators in ${\cal H}_{[p]}$, respectively, which are constructed out of the chiral fermions. In particular, these ${\cal J}_{\pm 1}$ operators obey the Heisenberg algebra
\be
\label{HA}
[{\cal J}_1, {\cal J}_{-1}] = 1, 
\ee
because they span the modes of a free chiral boson 
\be
\phi(z) = \phi_0 - i {\cal J}_0 \, {\rm ln}(z) + i \sum_{n \neq 0} {{\cal J}_n \over n}   \, z^{-n},
\ee
 where 
\be
\Psi(z) =  \sum_{r \in \mathbb Z + {1 \over 2}} \Psi_r \, z^{-r} = : e^{i \phi(z)}: \quad {\rm and} \qquad {\tilde \Psi} (z) =  \sum_{r \in \mathbb Z + {1 \over 2}} {\tilde \Psi}_r \, z^{-r} = : e^{-i \phi(z)}: 
\ee
is a \emph{single} chiral fermion and its conjugate such that  
\be
\Psi_{N(r + \rho_a)} = \psi^a_r, \qquad {\tilde \Psi}_{N(r - \rho_a)} = {\tilde \psi}^a_r,  \qquad \rho_a = {2a - N - 1 \over {2N}}.
\ee
We also have
\be
{\cal L}_0 = \sum_{r \in \mathbb Z + {1 \over 2}} r : \Psi_r {\tilde \Psi}_{-r}:.
\ee

Note that by using the commutator $[{\cal L}_0, {\cal J}_{\pm 1}] = {\cal J}_{\pm 1}$, we can, up to a constant factor of ${\Lambda}^{-Np^2 /2}$, write
\be
\label{TS limit}
Z_{\rm inst} (\Lambda', \hbar, \vec a) =  \langle p  | e^{{\Lambda' \over \hbar} {\cal J}_{1}}  \, e^{{\Lambda' \over \hbar}{\cal J}_{-1}} | p \rangle =  \langle {\rm coh}' | {\rm coh}' \rangle, 
\ee
where $\Lambda' = \Lambda^N$. Moreover, by comparing this with (\ref{q | q}), we find that we can interpret $| {\rm coh}' \rangle$ and $\langle {\rm coh}'|$ as a coherent state and its dual at the $z = 0$ and $z=\infty$ point on ${\cal C}_{\rm eff} = {\bf S}^2$.

\bigskip\noindent{\it A 5d AGT Correspondence for Pure $SU(N)$ in the Topological String Limit}

We are now ready to derive a 5d version of (\ref{TS limit}). To this end, first note that the 5d version of (\ref{q | q}) is just (\ref{q | q}) itself but at $\beta \neq 0$. As such, if $Z^{5d}_{\rm inst}(\Lambda', \hbar, \beta, \vec a)$ is the 5d Nekrasov instanton partition function at $\epsilon_1 = - \epsilon_2 = \hbar$, then, according to our discussion following (\ref{q | q}), we can write
\be
\label{Z5d}
Z^{5d}_{\rm inst} (\Lambda', \hbar, \beta, \vec a) =    \langle {\rm cir}'| {\rm cir}' \rangle, 
\ee
where $| {\rm cir}' \rangle$ is a coherent state that has a projection onto a circle $C_\beta$ of radius $\beta$ in ${\cal C}_{\rm eff} = {\bf S}^2$.

The explicit form of $| {\rm cir}' \rangle$ can be determined as follows. Firstly, recall that when a quantum system is confined to a space of infinitesimal size, its higher excited states would be decoupled; nevertheless, as we gradually increase the size of the space, we would start to observe these states. Now notice that $| {\rm coh}' \rangle = | p \rangle  + {\Lambda' \over \hbar}{\cal J}_{-1} | p \rangle + ({\Lambda' \over \hbar})^2{\cal J}_{-1} {\cal J}_{-1}| p \rangle + \dots$; the preceding statement and the last two paragraphs then mean that  $| {\rm cir}' \rangle = | p \rangle  + f_1(\Lambda', \hbar, \beta){\cal J}_{-1} | p \rangle + f_2(\Lambda', \hbar, \beta){\cal J}_{-2} | p \rangle +  f_{1,1}(\Lambda', \hbar, \beta){\cal J}_{-1} {\cal J}_{-1}| p \rangle + f_{2, 1}(\Lambda', \hbar, \beta){\cal J}_{-2} {\cal J}_{-1}| p \rangle + f_{1, 2}(\Lambda', \hbar, \beta){\cal J}_{-1} {\cal J}_{-2}| p \rangle+ f_{2,2}(\Lambda', \hbar, \beta){\cal J}_{-2} {\cal J}_{-2}| p \rangle + \dots$, where the $f$'s are constants depending on the indicated parameters. 

Secondly, notice that (i) $Z_{\rm BPS}$ in (\ref{5d BPS}) which underlies $Z^{5d}_{\rm inst}$ in (\ref{Z5d}) is invariant under the simultaneous rescalings $(\beta, \hbar, -\hbar, \vec a) \to ( \zeta \beta, \zeta^{-1} \hbar, - \zeta^{-1} \hbar, \zeta^{-1} \vec a)$, where $\zeta$ is some real constant; (ii)  $Z_{\rm inst}$ in (\ref{TS limit}) (which is just the $\beta \to 0$ limit of $Z^{5d}_{\rm inst}$) is invariant under the simultaneous rescalings $(\hbar, \Lambda') \to ( \zeta^{-1} \hbar, \zeta^{-1} \Lambda')$; and (iii) since the underlying worldvolume theory of the $N$ M5-branes on the LHS of (\ref{AGT-M-duality-AB}) is scale-invariant, it would mean that the physics ought to be invariant under the simultaneous rescalings $(\beta, \Lambda') \to (\zeta \beta, \zeta^{-1} \Lambda')$. Altogether, this means that the $f$'s should be invariant under the simultaneous rescalings $(\beta, \hbar, \Lambda') \to (\zeta \beta, \zeta^{-1} \hbar, \zeta^{-1} \Lambda')$. Thus,  the $f$'s should depend on $(\Lambda', \hbar, \beta)$ through  the rescaling-invariant combinations $\beta \hbar$ and $\beta \Lambda'$. 

Thirdly, notice that in (\ref{TS limit}), we can write the exponent ${({\Lambda' / \hbar}){\cal J}_{-1}}$ as $g(\Lambda', \hbar) \, {{\cal J}_{-m} \over -m}$, where $m= 1$ and $g$ is a constant depending on the indicated parameters. The preceding two paragraphs then mean that in order to arrive at  $| {\rm cir}' \rangle$, we ought to replace this exponent by $\sum_{n=1}^{\infty} f^n(\Lambda', \hbar, \beta) \, {{\cal J}_{-n} \over -n}$, where the $f^n$'s are distinct constants depending on  $\beta \hbar$ and $\beta \Lambda'$.

Fourthly, note that when $\beta \to 0$ whence only the ${\cal J}_{-1}$ mode acts on the vacuum state $| p \rangle$,   $|{\rm cir}' \rangle$ must reduce to $|{\rm coh}' \rangle$. On the other hand when $\beta \to \infty$, $|{\rm cir}' \rangle$  would be a state that has a projection onto a circle of infinite radius whence it would only have zero energy; in other words, $|{\rm cir}' \rangle$  must reduce to the vacuum state $| p \rangle$  when $\beta \to \infty$. 

Last but not least, notice that (\ref{coherent-A}) and (\ref{psi = delta}) mean that the $m^{\rm th}$ power of ${\Lambda'}$ ought to accompany the state of energy level $m$ in $|{\rm coh}' \rangle$ (which is indeed the case as can be seen from the expansion of $|{\rm coh}' \rangle$ four paragraphs earlier). In turn, this means that $f^n$ ought to depend on $(\beta \Lambda')^n$.  

The above five points and a little thought then lead one to conclude that 
\be
\label{Z5d-tilde}
\boxed{Z^{5d}_{\rm inst} (\Lambda', \hbar, \beta, \vec a) =    \langle {\rm cir}'| {\rm cir}' \rangle =  \langle p  | {\tilde \Gamma}_+ \, {\tilde \Gamma}_- | p \rangle}
\ee
where
\be
\label{tilde-gamma}
\boxed{{\tilde \Gamma}_-  = {\rm exp} \left(- \sum_{n=1}^\infty \, {(\beta \Lambda')^n \over 1 - q^n} \, {{\cal J}_{-n} \over n} \right) \qquad {\rm and} \qquad {\tilde \Gamma}_+ = {\rm exp} \left(\sum_{n=1}^\infty \, {(\beta \Lambda')^n \over 1 - q^{-n}} \, {{\cal J}_{n} \over n}\right)}
\ee
Here, $q = e^{\beta \hbar}$, and 
\be
\label{HA-5}
\boxed{[{\cal J}_m, {\cal J}_n] = m \delta_{m+n, 0}}
\ee

At any rate, note that by using the commutator $[{\cal L}_0, {\cal J}_{\pm m}] \sim {\cal J}_{\pm m}$, we can, up to a constant factor, also write (\ref{Z5d-tilde}) as
\be
\label{Z5d-NO}
{Z^{5d}_{\rm inst} (\Lambda', \hbar, \beta, \vec a) =   \langle p  | {\Gamma}_+ \, (\beta \Lambda')^{2 {\cal L}_0} { \Gamma}_- | p \rangle},
\ee
where
\be
\label{gamma-NO}
{{\Gamma}_-  = {\rm exp} \left( - \sum_{n=1}^\infty \, {1 \over 1 - q^n} \, {{\cal J}_{-n} \over n} \right) \qquad {\rm and} \qquad { \Gamma}_+ = {\rm exp} \left(\sum_{n=1}^\infty \, {1 \over 1 - q^{-n}} \, {{\cal J}_{n} \over n}\right)}.
\ee
The relations (\ref{Z5d-NO}) and (\ref{gamma-NO}) agrees with the results of Nekrasov-Okounkov in~\cite[$\S$7.2.3]{NO}, as expected. 

Comparing (\ref{Z5d-tilde})--(\ref{tilde-gamma}) with (\ref{TS limit}), it is clear that one can interpret $| {\rm cir}' \rangle$ as a state in an integrable module over some $q$-dependent affine algebra. One can ascertain this $q$-dependent affine algebra as follows. Firstly, recall that the (chiral) WZW model which underlies ${{\frak {su}}(N)}_{{\rm aff}, 1}$ on ${\cal C}_{\rm eff}$, can be regarded as a bosonic sigma model with worldsheet ${\cal C}_{\rm eff}$ and target an $SU(N)$ group manifold. Thus, $| {\rm coh}' \rangle$, which is defined over a point in ${\cal C}_{\rm eff}$, would be associated with a point in the space of all points into the target that is the $SU(N)$ group itself. Similarly, $| {\rm cir}' \rangle$, whose projection is onto a loop in ${\cal C}_{\rm eff}$, would be associated with a point in the space of all loops into the target that is the loop group of $SU(N)$. Via (\ref{Z5d-tilde}), this implies that the 5d $SU(N)$ theory underlying $Z^{5d}_{\rm inst}$ can also be interpreted as a 4d $LSU(N)$ theory, where $LSU(N)$ is the loop group of $SU(N)$. In turn, this means that we ought to replace ${\frak {su}}(N)$ with ${\frak {lsu}}(N)$, the Lie algebra of $LSU(N)$, throughout. Hence, since $| {\rm coh}' \rangle \in \widehat{{\frak {su}}(N)}_{{\rm aff}}$, we ought to have  
\be
\label{state double loop}
\boxed{| {\rm cir}' \rangle  \in {\widehat{{\frak {lsu}}(N)}}_{\rm aff}} 
\ee
where ${\widehat{{\frak {lsu}}(N)}}_{\rm aff}$ is an integrable module  over ${{\frak {lsu}}(N)}_{\rm aff}$, a universal central extension of the double loop algebra associated with $\frak{su}(N)$. This double loop algebra is the Lie algebra of smooth maps from the torus to $\frak{su}(N)$.\footnote{See~\cite[Part II, $\S$5]{Khesin} for more on this double loop algebra.}

Clearly, in arriving at the boxed relations (\ref{Z5d-tilde}), (\ref{tilde-gamma}), (\ref{HA-5}) and (\ref{state double loop}), we have just derived a 5d AGT correspondence for pure $SU(N)$ in the topological string limit!

\bigskip\noindent{\it The Appearance of a $q$-Deformed Affine ${\cal W}_N$-Algebra}

Before we end this section, let us make a final comment. We have hitherto studied only the $\epsilon_1 = - \epsilon_2$ topological string limit. What about the general case of arbitrary $\epsilon_{1,2}$? Since the pure case is just the case with a single adjoint hypermultiplet matter whose mass is infinitely large, as we shall see in the next section, the corresponding 5d pure Nekrasov instanton partition function can be expressed in terms of coherent states in an integrable module over a $q$-deformed affine ${\cal W}_N$-algebra.

\newsection{An M-Theoretic Derivation of a 5d AGT Correspondence}

\newsubsection{An M-Theoretic Derivation of a 4d AGT Correspondence with Matter: A Review}

Let us now review the M-theoretic derivation of a 4d AGT correspondence with matter in~\cite{4d AGT}. For brevity, we shall restrict our discussion to the conformal linear quiver of $n$ $SU(N)$ gauge groups.

In fig.~2(1) and fig.~2(2), we have the quiver diagram and its convenient schematic representation, respectively. As explained in~\cite{4d AGT}, the resulting effective correspondence between the original M-theory compactification and the I-brane ${\cal C}_{\rm eff} = {\bf S}^2$ in the dual compactification is as shown in fig.~2(3). Here, $\Sigma_t = {\bf S}^1 \times \mathbb I_t$, where $\beta$ is the radius of ${\bf S}^1$; the vertical planes represent the spatial part of M9-branes; $X^9\vert_{\epsilon_i} = \mathbb R^4\vert_{\epsilon_1, \epsilon_2} \times \mathbb R^5\vert_{\epsilon_3; x_{6,7}}$, where four of the spatial directions of the M5-branes are along $\mathbb R^4\vert_{\epsilon_1, \epsilon_2} \subset X^9\vert_{\epsilon_i}$; $\epsilon_3 = \epsilon_1 + \epsilon_2$;  $l = l_{\rm min}$ is the minimal instanton number of the underlying 4d $SU(N)$ theory along $\mathbb R^4\vert_{\epsilon_1, \epsilon_2}$; the $m_k$'s are the mass parameters associated with the flavor groups; $\Phi_{\vec \alpha, m_{i}, \vec \zeta}$ is an operator representing the shaded plane which transforms the theory with underlying $SU(N)$ Coulomb moduli $\vec \alpha$ to the adjacent theory with underlying $SU(N)$ Coulomb moduli $\vec \zeta$; the subscript `$\vec j_p$' is the highest weight that defines the primary operators $V^Q_{\vec j_p}$ and $V_{\vec j_p}$, where the superscript `$Q$' means that it also depends on $Q = \epsilon_3 /  \sqrt{\epsilon_1 \epsilon_2}$; $q_r = e^{2 \pi i \tau_r}$, where $\tau_r = 4 \pi i / g^2_r + \theta_r / 2 \pi$ is the complexified gauge coupling associated with the $r^{\rm th}$ gauge group; and the points where the $V_{\vec j_p}$'s are inserted are $z = 1, q_1, q_1q_2, \dots, q_1q_2\dots q_n$. 

\begin{figure}
  \centering
    \includegraphics[width=0.8\textwidth]{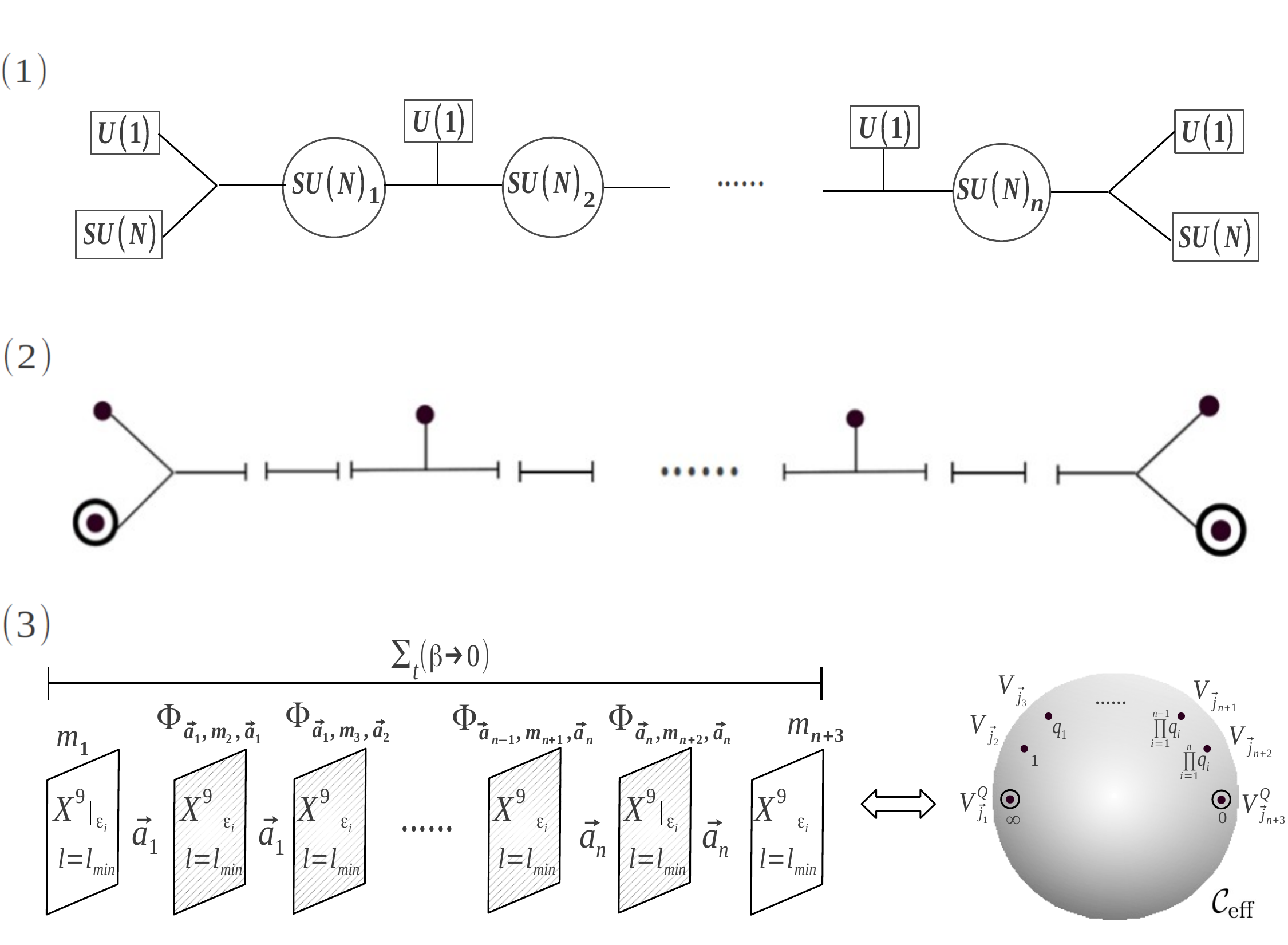}
  \caption{(1) The linear quiver diagram; (2) its schematic representation; (3) the resulting effective correspondence}
\end{figure}

Clearly, the operator $\Phi_{\vec \alpha, m_{i}, \vec \zeta}$ also transforms the space (\ref{BPS-AGT-AB}) of BPS states of the theory with parameter $\vec \alpha$ to that of the adjacent theory with parameter $ \vec \zeta$. Thus, we can also describe $\Phi$ as the following map:
\be
\label{Phi-map}
\Phi_{\vec \alpha, m_{i}, \vec \zeta}: {\cal H}_{\vec \alpha} \to {\cal H}_{\vec \zeta}, \quad {\rm where} \quad {\cal H}_{\vec a_p} = \bigoplus_{l} ~{\rm IH}^\ast_{U(1)^2 \times T} \, {\cal U}({\cal M}_{SU(N), l}) \otimes \mathbb C(\epsilon_1, \epsilon_2, \vec a_p),
\ee
and ${\cal H}_{\vec a_p}$ is the space of BPS states of the theory with parameter $\vec a_p$.

At any rate, in the case of a conformal linear quiver of $n$ $SU(N)$ gauge groups, the expression (\ref{Z_inst-5d}) for the 4d Nekrasov instanton partition function ought to be replaced by
\be
\label{Z-lin}
Z^{\rm lin}_{\rm inst}({\bf q}, \epsilon_1, \epsilon_2, {\vec {\bf a}}, {\bf m}) =  \sum_{l_1, l_2 \dots, l_n} q^{l_1}_1 q^{l_2}_2 \cdots q^{l_n}_n \, Z^{\rm lin}_{{\rm BPS}, l_1, l_2, \dots, l_n} (\epsilon_1, \epsilon_2, \vec {\bf a}, {\bf m}, \beta \to 0), 
\ee 
where $l_i$ is the instanton number associated with the $SU(N)_i$ gauge group, and $Z^{\rm lin}_{{\rm BPS}, l_1, l_2, \dots, l_n}$ is the partition function of the aforementioned BPS states associated with the left diagram in fig.~2(3). This partition function can be viewed as a sum over BPS states that propagate from the rightmost to the leftmost end of the diagram which undergo transformations of the kind described in (\ref{Phi-map}) due to the presence of the shaded planes; in other words, one can also write 
\be
\label{correlation of Phi's}
\hspace{-0.2cm}Z^{\rm lin}_{\rm inst}({\bf q}, \epsilon_1, \epsilon_2, {\vec {\bf a}}, {\bf m}) = {}_{m_1}\langle \emptyset \vert  \Phi_{\vec a_1, m_{2}, \vec a_{1}} \, q^{{\bf l}_1}_1 \, \Phi_{\vec a_1, m_{3}, \vec a_{2}} \, q^{{\bf l}_2}_2 \cdots  \Phi_{\vec a_{n-1}, m_{n+1}, \vec a_{n}} \, q^{{\bf l}_n}_n \,\Phi_{\vec a_{n}, m_{n+2}, \vec a_{n}} \vert \emptyset \rangle_{m_{n+3}},
\ee   
where ${}_{m_1}\langle \emptyset \vert$ and $ \vert \emptyset \rangle_{m_{n+3}}$ are the minimum energy BPS states at the leftmost and rightmost end of the diagram that are associated with $m_1$ and $m_{n+3}$, respectively, while ${\bf l}_i$ is an instanton number operator whose eigenvalue is the instanton number $l_i$ associated with the BPS states.

Note at this point that the duality relation in (\ref{AGT-duality-A}), the discussion following it, and the map (\ref{Phi-map}), also mean that 
 \be
 \label{phi-tx}
\Phi_{\vec \alpha, m_{i}, \vec \zeta}: {\cal V}_{{\bf j}(\vec \alpha)} \to {\cal V}_{{\bf j}(\vec \zeta)}, 
\ee
where $ {\cal V}_{{\bf j}(\vec a_p)}$ is a Verma module over an affine ${\cal W}_N$-algebra of central charge and highest weight 
\be
\label{cc-linear quiver}
{c = (N-1)  + (N^3 - N) {\epsilon_3^2 \over \epsilon_1 \epsilon_2}} \quad {\rm and} \quad {\bf j}(\vec a_p) = {-i\vec a_p \over {\sqrt{\epsilon_1 \epsilon_2}}}  + i Q \vec\rho,
\ee 
with $\vec \rho$ being the Weyl vector of $\frak {su}(N)$. Consequently, $\Phi$ can also be interpreted as a primary vertex operator $V$ acting on $\cal V$; this underlies the correspondence between $\Phi_{\vec \alpha, m_{i}, \vec \zeta}$ and $V_{\vec j_{i}}$ in fig.~2(3).  Similarly, the duality relation in  (\ref{AGT-duality-A}), and the discussion following it, underlie the correspondence between ${}_{m_1}\langle \emptyset \vert$ and $ \vert \emptyset \rangle_{m_{n+3}}$ and $\langle V^Q_{\vec j_1}\vert$ and $\vert V^Q_{\vec j_{n+3}}\rangle$ in fig.~2(3). 

Hence, the correspondence depicted in fig.~2(3), and our explanations in the last three paragraphs, mean that we can write 
\be
\label{AGT-lin}
\hspace{-0.0cm}{Z^{\rm lin}_{\rm inst}({\bf q}, \epsilon_1, \epsilon_2, {\vec {\bf a}}, {\bf m}) = Z^{\rm lin}({\bf q}, \epsilon_i, {\bf m}) \cdot \left\langle V^Q_{\vec j_1} (\infty) V_{\vec j_2} (1) V_{\vec j_3} (q_1) \dots  V_{\vec j_{n+2}} (q_1q_2 \dots q_{n}) V^Q_{\vec j_{n+3}} (0) \right\rangle_{{\bf S}^2}}.
\ee
The independence of the factor $Z^{\rm lin}$ on $\vec {\bf a}$ is because the $\vec a_p$'s have already been ``contracted'' in the correlation function -- see the RHS of (\ref{correlation of Phi's}).   

According to (\ref{cc-linear quiver}), the fact that the $a_p$'s and the $m_k$'s have the same dimension, and the fact that $V^Q_{j_1}$ and $V^Q_{j_{n+3}}$ ought to depend on $m_1$ and $m_{n+3}$, respectively, one can conclude that 
\be
\label{vecj-1}
{\vec j_1 = {-i{\vec m}_1  \over {\sqrt{\epsilon_1 \epsilon_2}}}  + {i \vec\rho \, \epsilon_3  \over \sqrt{\epsilon_1 \epsilon_2}} \quad {\rm and} \quad \vec j_{n+3} = {-i \vec m_{n+3} \over {\sqrt{\epsilon_1 \epsilon_2}}}  + {i\vec\rho \, \epsilon_3  \over \sqrt{\epsilon_1 \epsilon_2}}},
\ee
where the $N-1$ component vectors $\vec m_1$ and $\vec m_{n+3}$ depend on $m_1$ and $m_{n+3}$. 

Similarly, one can conclude, after noting that $Q$ vanishes where the $ V_{\vec j_u}$ operators are inserted~\cite{4d AGT}, that  
\be
\label{vecj-2}
{\vec j_u = {-i{\vec m}_u  \over {\sqrt{\epsilon_1 \epsilon_2}}} \quad {\rm for} \quad u = 2, 3, \dots, n+2},
\ee
where the $N-1$ component vector $\vec m_i$ depends on $m_i$. 

\begin{figure}
  \centering
    \includegraphics[width=0.3\textwidth]{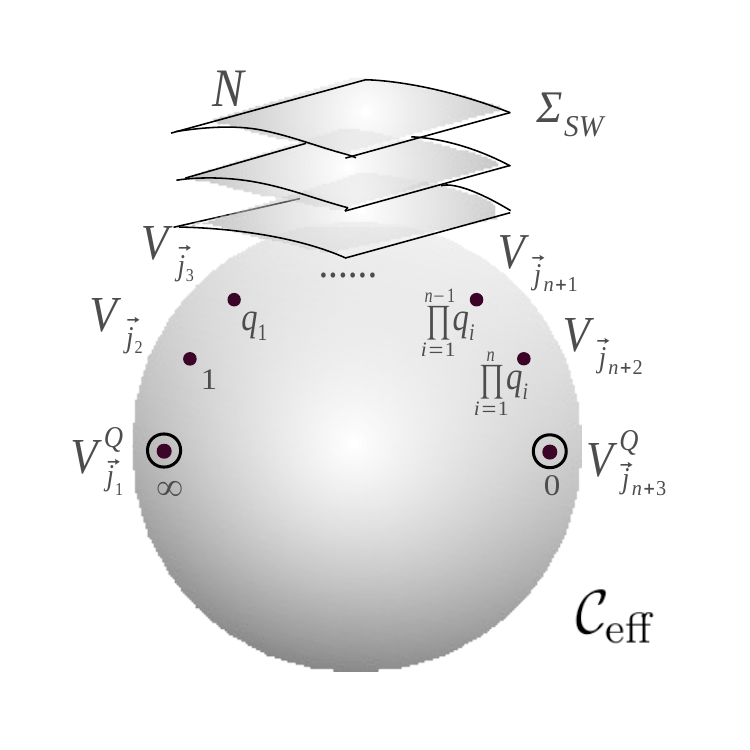}
  \caption{${\cal C}_{\rm eff}$ and its $N$-fold cover $\Sigma_{SW}$ with primary operators inserted at the $n+3$ punctures $z= \infty, 1, q_1, q_1q_2, \dots, q_1q_2 \dots q_n, 0$.}
\end{figure}

Last but not least, note that like in $\S$2.1, we effectively have $N$ D6-branes and $1$ D4-brane wrapping ${\cal C}_{\rm eff}$, i.e., we effectively have an $N \times 1 = N$-fold cover $\Sigma_{SW}$ of ${\cal C}_{\rm eff}$. This is depicted in fig.~3. Incidentally, $\Sigma_{SW}$ is also the Seiberg-Witten curve which underlies  $Z^{\rm lin}_{\rm inst}({\bf q}, \epsilon_1, \epsilon_2, {\vec {\bf a}}, {\bf m})$! In fact, $\Sigma_{SW}$ can be described in terms of the algebraic relation~\cite{N=2} 
\be
\label{SW-matter}
\Sigma_{SW}: \lambda^N + \sum_{k=2}^{N} \lambda^{N-k} \phi_k(z) = 0,
\ee 
where $\lambda = y  dz /z$ (for some complex variable $y$) is a section of $T^\ast {\cal C}_{\rm eff}$, and the $\phi_k(z)$'s are $(k,0)$-holomorphic differentials on ${\cal C}_{\rm eff}$  with poles at the punctures $z = \infty, 1, q_1, q_1q_2, \dots, \newline q_1q_2 \dots q_n$ that are determined by the matter content of the 4d theory. In particular, near the puncture $z = z_s$, we have 
\be
\phi_2(z) \sim {u^{(2)}_s dz^2 \over (z - z_s)^2}, 
\ee
and from the correspondence between $\phi_2(z)$ and the holomorphic stress tensor $W^{(2)}(z)$ (established in $\S$2.1, which thus applies here too), we have 
\be
W^{(2)} (z) {\mathbb V}_{\vec j_s} (z_s) \sim {u^{(2)}_s  \over (z - z_s)^2}  {\mathbb V}_{\vec j_s} (z_s),
\ee
where ${\mathbb V}_{\vec j_s} (z_s)$ can be $V^Q_{\vec j_s} (z_s)$ or $V_{\vec j_s} (z_s)$.  In other words, the conformal dimension of the primary operator $ {\mathbb V}_{\vec j_s} (z_s)$ is equal to $u^{(2)}_s$, i.e., we have
\be
\label{weight-lin}
{ {{\vec j_s}^2 \over 2} -   {i \epsilon_3 \,  \over \sqrt{\epsilon_1 \epsilon_2}} \, \vec j_s \cdot  \vec\rho = u^{(2)}_s, \quad {\rm where} \quad s = 1, 2, \dots, n+3},
\ee
from which we can ascertain the explicit form of the mass vectors $\vec m_s$ in (\ref{vecj-1}) and (\ref{vecj-2}). 

In arriving at the relations (\ref{Phi-map}), (\ref{phi-tx}), (\ref{cc-linear quiver}), (\ref{AGT-lin}), (\ref{vecj-1}), (\ref{vecj-2}) and (\ref{weight-lin}), we have just derived the 4d AGT correspondence for a conformal linear quiver of $n$ $SU(N)$ gauge groups.

\newsubsection{An M-Theoretic Derivation of a 5d AGT Correspondence With Matter}

Let us now proceed to derive a 5d analog of the 4d AGT correspondence with matter. For brevity, and to make contact with relevant results in the mathematical literature, we shall restrict ourselves to an $SU(N)$ theory with $N_f = 2N$ fundamental matter, i.e., $n=1$ in the last subsection. 

\bigskip\noindent{\it A 5d AGT Correspondence for $U(1)$ with $N_f = 2$ Fundamental Matter}

It would be useful to first consider the $N=1$ case, i.e., the $U(1)$ theory with $N_f = 2$ fundamental matter. To this end, first notice that we can also write (\ref{TS limit}) as the three-point correlation function on ${\bf S}^2$:
\be
\label{Z4d-massive limit}
Z_{\rm inst} (\Lambda', \hbar, \vec a) =    \langle V_0 (\infty) \,  V_{\Gamma} (1) \, V_0 (0) \rangle_{{\bf S}^2}, 
\ee
where the vertex operators $V_0 (\infty)$ and $V_0 (0)$ represent the vacuum states $\langle 0 |$ and $| 0 \rangle$, respectively, while the operator
\be
\label{V-gamma}
V_{\Gamma} (1) =   {\rm exp} \left({ \Lambda' \over \hbar} {\cal J}_{1} \right) {\rm exp} \left( { \Lambda' \over \hbar} {\cal J}_{-1} \right). 
\ee
Here, $\Lambda'$ is the energy scale, and ${\cal J}_{\pm 1}$ are operators that obey the Heisenberg algebra (\ref{HA}). As a pure $U(1)$ theory can also be interpreted as the $m \to \infty$, $e^{2 \pi i \tau} \to 0$ limit of a $U(1)$ theory with an adjoint hypermultiplet matter of mass $m$ and complexified gauge coupling $\tau$, the analysis in~\cite[$\S$6.3]{4d AGT} would mean that we can regard the $V$'s here as \emph{primary} vertex operators.\footnote{This can be understood as follows. Firstly, note that the case of a $U(1)$ theory with an adjoint hypermultiplet matter corresponds to~\cite[fig.~8]{4d AGT} with $N = n=1$. In the limit $e^{2 \pi i \tau} = e^{ -{8 \pi^2 \over g^2} + i \theta} \to 0$, i.e., when ${1 \over g^2} \to \infty$,  the I-brane ${\cal C}_{\rm eff}$ in \emph{loc.~cit.} -- after an allowed conformal transformation to bring the two infinitely-long tubes (of length $1 \over g^2$) back to the origin whilst replacing the resulting holes with vertex operators -- would be an ${\bf S}^2$ with vertex operators inserted at $z= 0$, $1$ and $\infty$. Then, as explained in  \emph{loc.~cit.}, the original vertex operator at $z=1$ is primary, and since  the vertex operators at $z =0, \infty$ are supposed to represent far away states whose excitations have decayed to zero, they must also be primary. \label{massive limit}}  

Similarly, (\ref{Z5d-tilde}) can also be expressed as the three-point correlation function on ${\bf S}^2$:
\be
\label{Z5d-tilde-massive limit}
 Z^{5d}_{\rm inst} (\Lambda', \hbar, \beta, \vec a) =   \langle V_0 (\infty) \,  {\tilde V}_{\Gamma} (1) \, V_0 (0) \rangle_{{\bf S}^2}, 
\ee
where the vertex operator
\be
\label{V-tilde-gamma}
{\tilde V}_{\Gamma} (1)  = {\rm exp} \left(\sum_{n=1}^\infty {1\over n} {(\beta \Lambda')^n \over 1 - q^{-n}} \, {{\cal J}_{n}} \right) {\rm exp} \left( - \sum_{n=1}^\infty { 1 \over n} {(\beta \Lambda')^n \over 1 - q^n} \, {{\cal J}_{-n}} \right), 
\ee
$q = e^{\beta \hbar}$, $\beta$ is the radius of the 5th dimension circle, and ${\cal J}_{\pm m}$ are operators that obey the Heisenberg algebra (\ref{HA-5}). The last paragraph then means that we can regard ${\tilde V}_{\Gamma}$ as a 5d analog of the 4d primary vertex operator $V_\Gamma$.  In particular, all 5d analogs of 4d primary vertex operators ought to take a form similar to ${\tilde V}_{\Gamma}$ in (\ref{V-tilde-gamma}), where one recalls that $\beta \Lambda'$ is a rescaling-invariant combination of parameters. 

Second, note that for general values of $\epsilon_{1,2}$, according to our discussion in $\S$2.1 leading up to (\ref{AGT-AB-chiral CFT}), and the fact that the smallest connected subgroup of $U(1)$ is itself, the relevant 2d CFT along ${\cal C}_{\rm eff} = {\bf S}^2$ is (the chiral half of) a $G/G$ WZW model, where $G = U(1)$. As explained in $\S$2.2, any state in the 5d case has a projection onto a circle $C_\beta$ of radius $\beta$ in ${\cal C}_{\rm eff}$. Notably, an arbitrary state $| \lambda \rangle$ in the 5d case ought to be represented by a corresponding gauge-invariant loop operator in the $G/G$ WZW model. Now cut out two circles $C_\beta$ at the poles of ${\cal C}_{\rm eff}$ and wrap a Wilson loop ${\rm Tr}_{R_\nu} \, {\rm exp} \oint_{C_\beta} A$ in some representation $R_\nu$ around them; this should correspond to the inner product $\langle \lambda | \lambda \rangle$. As the $G/G$ WZW model is topological, one can also put the theory on the annulus, where the monodromies of the restrictions of the $G$-connection $A$ coincide on the two circle boundaries. The inner product $\langle \lambda | \lambda \rangle$ can then be computed as the path integral over this annulus integrated over the space of all boundary monodromies weighted by ${\rm Tr}_{R_\nu} \, {\rm exp} \oint_{C_\beta} A$.  In turn, according to~\cite[$\S$2.4]{NG}, $\langle \lambda | \lambda \rangle$  would be expressed in terms of Macdonald polynomials; in other words, there is a correspondence between $|\lambda \rangle$ and Macdonald polynomials.\footnote{In \emph{loc.~cit.}, the relevant theory actually has action $S = S_{G/G} + \int {\rm Tr} \, (\dots)$,  where $S_{G/G}$ is the action of the $G/G$ WZW model. Nevertheless, in the computation of the corresponding path integral (which gives rise to Macdonald polynomials), the contributions coming from the additional term $\int {\rm Tr} \, (\dots)$ just serve to correct $\langle \lambda | \lambda \rangle$; since a Macdonald polynomial can always be written as a sum of other Macdonald polynomials, $\langle \lambda | \lambda \rangle$ would also be expressed in terms of Macdonald polynomials, as claimed.}  Consequently, according to~\cite[$\S$2.3]{Awata}, this means that we can express 
\be
\label{lambda states}
| \lambda \rangle = a_{-\lambda _1} a_{-\lambda_2} \dots | \emptyset \rangle, \quad  a_{\lambda_i > 0} | \emptyset \rangle = 0,  
\ee 
where the $\lambda_i$'s define a partition $\lambda = (\lambda_1, \lambda_2, \cdots)$ associated with the corresponding Macdonald polynomial, $| \emptyset \rangle$ is the vacuum state, while the $a_i$'s are operators obeying a deformed Heisenberg algebra: 
\be
[a_m, a_n] = m {1- {\tilde q}^{|m|} \over 1- {\tilde t}^{|m|}} \delta_{m+n, 0},
\ee
with $\tilde q$ and $\tilde t$ being independent parameters. 

Third, from (\ref{Phi-map}), \cite[Theorem 1]{CO}, and  (\ref{phi-tx}), we find that in the $N=1$ case at hand, the 4d Nekrasov instanton partition function (\ref{AGT-lin}) would (up to some constant factor) be given by
\be
\label{AGT-lin-U(1)}
\hspace{-0.0cm}{Z^{\rm lin}_{{\rm inst}, \, U(1)}({q_1}, \epsilon_1, \epsilon_2, {\vec m}) =  \langle 0 | V_{{\tilde m}_1} (1) V_{{\tilde m}_2} (q_1) | 0 \rangle_{{\bf S}^2}}.
\ee
Here, $q_1 = e^{2 \pi i \tau}$, where $\tau$ is the complexified gauge coupling; ${\vec m} = (m_1, m_2)$ are the $N_f = 2$ fundamental matter masses; $\tilde m_{1, 2}$ are independent linear combinations of $m_{1,2}$; the vacuum states $| 0 \rangle$ and $\langle 0 |$ are represented by the $V^Q$ primary vertex operators inserted at $z = 0$ and $\infty$, respectively; and the primary vertex operators
\be
\label{4d vertex operators-U(1)}
V_{{\tilde m}_k}(z) =  {\rm exp} \left[ -{i \tilde m_k \over \sqrt{\epsilon_1 \epsilon_2}}  \varphi_+(z) \right] \, {\rm exp} \left[- {i \tilde m_k  - (\epsilon_1 + \epsilon_2) \over \sqrt{\epsilon_1 \epsilon_2}}  \varphi_-(z) \right],
\ee 
where $\varphi(z)$ is a free-boson field with OPE $\varphi(z) \varphi(0) \sim - {\rm ln} \, z$, and $\varphi(z) = \varphi_+(z) + \varphi_-(z)$ is its decomposition into positive and negative modes, i.e.,
\be
\label{varphi}
\varphi_-(z) = -\sum_{n > 0}  {b_{-n} \over  n} \, z^n \qquad {\rm and} \qquad \varphi_+(z) = \sum_{n > 0}  {b_{n} \over n} \, z^{-n},
\ee
with 
\be
\label{b-commutator}
[b_m, b_n] = f(\epsilon_1, \epsilon_2) \, m  \delta_{m+n, 0} \qquad \textrm{where} \qquad  f_{\epsilon_1 =- \epsilon_2 } \to 1.\footnote{Recall that when $\epsilon_3 = \epsilon_1 + \epsilon_2 = 0$, there is no Omega-deformation on the RHS of (\ref{AGT-M-duality-AB}) whence the aforestated affine algebra along $\cal C$ reverts to the undeformed Heisenberg algebra ${{{\frak {u}}(1)}_{{\rm aff}, 1}}$. Hence, we necessarily have $f_{\epsilon_1 =- \epsilon_2 } \to 1$.}
\ee
In particular, in the limit $\beta \to 0$, the 5d analog of $V_{\tilde m_k}$ should reduce to (\ref{4d vertex operators-U(1)}) with the accompanying relations (\ref{varphi}) and (\ref{b-commutator}). 

In sum, the above three points mean that we can express the 5d Nekrasov instanton partition function for $U(1)$ with $N_f = 2$ fundamental matter as
\be
\label{Z5d-U(1)}
\boxed{{Z^{\rm lin, \, 5d}_{{\rm inst}, \, U(1)}({q_1}, \epsilon_1, \epsilon_2, \beta, {\vec m}) =  \langle \emptyset |  {\Phi}^w_v (z_1) \Phi^v_u (z_2)  | \emptyset \rangle_{{\bf S}^2}}}
\ee
where the vacuum states $| \emptyset \rangle$ and $\langle \emptyset |$ represent the 5d analogs of the 4d  primary vertex operators $V^Q$ inserted at $z = 0$ and $\infty$, respectively,
\be
\label{VO-U(1)}
\boxed{ \Phi^c_d (z) =   {\rm exp} \left(  \sum_{n > 0} {1 \over n} {c^{-n} -  d^{-n} \over  1 - q^{-n}} \, {a}_{n} \, z^{-n} \right) {\rm exp} \left( - \sum_{n > 0} {1 \over n} {c^n - (t / q)^n d^n \over  1 - q^n} \, {a}_{-n} \, z^n \right)}  
\ee
and the rescaling-invariant parameters
\be
\label{variables-U(1)}
\boxed{w = e^{- \beta m_1},  \quad v = e^{-2\beta (m_1 + m_2)}, \quad u = e^{- \beta m_2}, \quad q = e^{- i \beta \sqrt{\epsilon_1 \epsilon_2}}, \quad t = e^{- i \beta ( \epsilon_1 + \epsilon_2 + \sqrt{\epsilon_1 \epsilon_2})}} 
\ee
while
\be
\boxed{[{a}_m, {a}_n] = m {1- { q}^{|m|} \over 1- { t}^{|m|}} \delta_{m+n, 0}, \quad a_{m > 0} | \emptyset \rangle = 0}
\label{he}
\ee
Clearly, in arriving at the boxed relations (\ref{Z5d-U(1)}), (\ref{VO-U(1)}), (\ref{variables-U(1)}) and (\ref{he}), we have just derived a 5d AGT correspondence for $U(1)$ with $N_f = 2$ fundamental matter!

\bigskip\noindent{\it The Ding-Iohara Algebra}

In fact, (\ref{Z5d-U(1)}) and (\ref{VO-U(1)}) (bearing in mind (\ref{he})) coincide with \cite[eqns.~(2.7) and (2.6)]{Awata}. As such, $\Phi^c_d (z)$ must be a vertex operator for a level one representation of the Ding-Iohara algebra~\cite{DI}; in other words, $Z^{\rm lin, \, 5d}_{{\rm inst}, \, U(1)}$ is actually expressed in terms of  a level one module over the Ding-Iohara algebra.  This important observation will soon allow us to connect the $N > 1$ case with a $q$-deformed affine ${\cal W}_N$-algebra.

\bigskip\noindent{\it A 5d AGT Correspondence for $SU(N)$ with $N_f = 2N$ Fundamental Matter}

Let us now proceed to consider an $SU(N)$ theory with $N_f = 2N$ fundamental matter. Recall from fig.~3 that for general $N$, we effectively have $N$ D6-branes and $1$ D4-brane wrapping ${\cal C}_{\rm eff} = {\bf S}^2$ -- that is, we effectively have an $N \times 1 = N$-fold cover of ${\cal C}_{\rm eff}$. This means that we ought to make the following replacements in (\ref{Z5d-U(1)}) to get the corresponding result for the $SU(N)$ case. 

Firstly, the vacuum states $\langle \emptyset |$ and $| \emptyset \rangle$ ought to be replaced by their $N$-tensor product:
\be 
\label{Z5d-SU(N)-matter-vacuum}
\boxed{ \langle {\bf 0} | = \langle \emptyset |^{\otimes N} \quad {\rm and} \quad  | {\bf 0} \rangle =  | \emptyset \rangle^{\otimes N}}
\ee

Secondly, the vertex operators ${\Phi}^w_v$ and $ \Phi^v_u$ ought to likewise be replaced by their $N$-tensor product
\be
\label{Z5d-SU(N)-matter-vertex operator}
\boxed{{\Phi}^{\bf w}_{\bf v} (z_1) = {\Phi}^{w_N}_{v_N} \cdots {\Phi}^{w_1}_{v_1} (z_1)  \qquad {\rm and} \qquad  \Phi^{\bf v}_{\bf u} (z_2) =   \Phi^{v_1}_{u_1} \cdots \Phi^{v_N}_{u_N} (z_2)}
\ee
where ${\bf c} = (c_1, c_2, \dots, c_N)$. 

In sum, we can express the 5d Nekrasov instanton partition function for $SU(N)$ with $N_f = 2N$ fundamental matter as
\be
\label{Z5d-SU(N)}
\boxed{{Z^{\rm lin, \, 5d}_{{\rm inst}, \, SU(N)}({q_1}, \epsilon_1, \epsilon_2, \beta, {\vec a}, {\vec m}) =  \langle {\bf 0} |  {\Phi}^{\bf w}_{\bf v} (z_1) \Phi^{\bf v}_{\bf u} (z_2)  | {\bf 0} \rangle_{{\bf S}^2}}}
\ee
and if the vertex operator ${\Phi}^{ c}_{d}$  defines a map ${\cal F}_{d} \to {\cal F}_{c}$ from one Fock space associated with parameter $d$ to another associated with parameter $c$, then
\be
\label{Z5d-SU(N)-matter-vertex operator-map}
{{\Phi}^{\bf c}_{\bf d}: {\cal F}_{d_1} \otimes {\cal F}_{d_2} \otimes \cdots \otimes {\cal F}_{d_N} \longrightarrow {\cal F}_{c_1} \otimes {\cal F}_{c_2} \otimes \cdots \otimes {\cal F}_{c_N}}.
\ee

Note that in the $U(1)$ case, the dependence of ${\Phi}^{ c}_{d}$ on the Coulomb modulus actually drops out where it is replaced by a linear combination of the masses $m_{1,2}$ which appears in the parameter $v$ in (\ref{variables-U(1)}). In other words, in the $SU(N)$ case, ${\bf v} = (v_1, \dots, v_N)$ would be associated with the Coulomb moduli ${\vec a} = (a_1, \dots, a_N)$ (where $\sum_i a_i = 0$), while ${\bf w} = (w_1, \dots, w_N)$ and ${\bf u} = (u_1, \dots, u_N)$ would continue to be associated with the $2N$ masses ${\vec m} = (m_1, m_2, \dots, m_{2N})$. As the parameters ${\bf w}, {\bf v}, {\bf u}$ must reduce to $w, v, u$ in (\ref{variables-U(1)}) when $N=1$, we conclude that 
\be
\label{Z5d-SU(N)-variables}
\boxed{w_i = e^{- \beta m_i},  \quad v_i = e^{-\beta a_i}, \quad u_i = e^{- \beta m_{N+i}}}
\ee

As mentioned, ${\Phi}^{ c}_{d}$ is a vertex operator of a level one representation of the Ding-Iohara algebra -- that is, ${\cal F}_{c, d}$ is a level one module over the Ding-Iohara algebra. As such, (\ref{Z5d-SU(N)-matter-vertex operator-map}) means that ${\Phi}^{\bf c}_{\bf d}$ is a vertex operator of a level $N$ representation of the Ding-Iohara algebra~\cite{Awata}. In turn, this means that 
\be
\label{q-W}
\boxed{{\Phi}^{\bf c}_{\bf d}:  \, \widehat {{\cal W}^q_N} \to \widehat {{\cal W}^q_N}}
\ee
where $\widehat {{\cal W}^q_N}$ is a Verma module over ${\cal W}^q_N$, a $q$-deformed affine ${\cal W}_N$-algebra~\cite{FHSSY}.\footnote{In \emph{loc.~cit.}, it was actually shown that the aforementioned level $N$ representation of the Ding-Iohara algebra is realized by a representation of ${\cal W}^q_N \otimes {\frak u}(1)_{\rm aff}$. However, as explained in~\cite[$\S$5.2]{4d AGT}, because the $U(1)$ gauge field on the D4-brane in (\ref{equivalent IIA system 1 - AGT}) --  unlike the $SU(N)$ gauge field on the D6-branes which it intersects -- is dynamical,  one has to reduce away in the I-brane system the $U(1)$ WZW model associated with the D4-brane. This is equivalent to reducing away the ${\frak u}(1)_{\rm aff}$ factor, whence (\ref{q-W}) is indeed consistent.}

Clearly, in arriving at the boxed relations (\ref{Z5d-SU(N)-matter-vacuum}), (\ref{Z5d-SU(N)-matter-vertex operator}), (\ref{Z5d-SU(N)}),  (\ref{Z5d-SU(N)-variables}) and (\ref{q-W}), we have just derived a 5d AGT correspondence for $SU(N)$ with $N_f = 2N$ fundamental matter! This serves as a purely physical M-theoretic proof of~\cite[Conjecture 3.13]{Awata}, and also lends further weight to the interesting analysis in~\cite{Ito} (which is based on the assumption that the correspondence holds).

\newsubsection{An M-Theoretic Derivation of a 5d Pure AGT Correspondence} 

Let us now proceed to derive a 5d analog of the 4d pure AGT correspondence. Again, it would be useful to start with the $U(1)$ case. 

\bigskip\noindent{\it A 5d AGT Correspondence for Pure $U(1)$}

 As mentioned in the previous subsection, since a pure $U(1)$ theory can also be interpreted as the $m \to \infty$, $e^{2 \pi i \tau'} \to 0$ limit of a $U(1)$ theory with an adjoint hypermultiplet matter of mass $m$ and complexified gauge coupling $\tau'$, where $m e^{   2\pi i \tau' } = \Lambda$ remains fixed, the analysis in~\cite[$\S$6.3]{4d AGT} and footnote~\ref{massive limit} would mean that the 5d Nekrasov instanton partition function for pure $U(1)$ can be expressed as
\be
\label{1-point-U(1)}
{Z^{{\rm pure}, \, 5d}_{{\rm inst}, \, U(1)} (\epsilon_1, \epsilon_2, \beta, \Lambda)  =   \langle \emptyset |  {\Phi}_{m \to \infty} (1) | \emptyset \rangle_{{\bf S}^2}},
\ee 
where ${\Phi}_{m \to \infty}(1)$ is the 5d analog of the 4d primary vertex operator $V_{m}(1)$ in (\ref{4d vertex operators-U(1)}) in the $m \to \infty$ limit. In other words, according to our derivation of (\ref{VO-U(1)})--(\ref{he}) from (\ref{4d vertex operators-U(1)}) etc., we find that 
\be
{{\Phi}_{m \to \infty} (1) = {\rm exp} \left(  \sum_{n > 0} {1 \over n} {c^{-n} -  d^{-n} \over  1 - q^{-n}} \, {a}_{n} \right) {\rm exp} \left( - \sum_{n > 0} {1 \over n} {c^n - (t / q)^n d^n \over  1 - q^n} \, {a}_{-n} \right)},
\ee
so we can write
\be
\label{Z5d-pure-U(1)-initial}
Z^{{\rm pure}, \, 5d}_{{\rm inst}, \, U(1)} (\epsilon_1, \epsilon_2, \beta, \Lambda)  = \langle G_{U(1)} | {\mathscr G}_{U(1)} | G_{U(1)} \rangle, 
\ee
where
\be
{\mathscr G}_{U(1)} = {\rm exp} \left(  \sum_{n >0} {1 \over n} {d^n(t^n - q^n) \over q^n(1 - q^n)} {a}_{-n} \right), 
\ee
\be
\label{Z5d-G-pure U(1)-initial}
{| G_{U(1)} \rangle =  {\rm exp} \left( -  \sum_{n > 0} {1 \over n} {c^n -  d^n \over  1 - q^n} \, {a}_{-n} \right) | \emptyset \rangle,  \quad  \langle G_{U(1)} | = \langle \emptyset | {\rm exp} \left(   \sum_{n > 0} {1 \over n} {c^{-n} -  d^{-n} \over  1 - q^{-n}} \, {a}_{n} \right)},
\ee
and the rescaling-invariant parameters $c = e^{- \beta ({m \to \infty})}$,  $d = e^{-2\beta  ({m \to \infty})}$, $q = e^{- i \beta \sqrt{\epsilon_1 \epsilon_2}}$, $t = e^{- i \beta ( \epsilon_1 + \epsilon_2 + \sqrt{\epsilon_1 \epsilon_2})}$, while
${[{a}_p, {a}_n] = p {1- { q}^{|p|} \over 1- { t}^{|p|}} \delta_{p+n, 0}}$.

Nonetheless, $d \to 0$ so $\mathscr G_{U(1)} \to 1$ whence we can write (\ref{Z5d-pure-U(1)-initial}) as
\be
\label{Z5d-pure U(1)}
\boxed{Z^{{\rm pure}, \, 5d}_{{\rm inst}, \, U(1)} (\epsilon_1, \epsilon_2, \beta, \Lambda)  = \langle G_{U(1)}  | G_{U(1)} \rangle}
\ee
Also, note that (i) for $n > 0$, we have $(c^{-n} - c^n) = (d^{-n} -d^n) \to \infty$, i.e., $(c^{-n} - d^{-n}) = (c^n - d^n)$; (ii) since $c$ and $d$ are rescaling-invariant, so must $(c^n - d^n)$; (iii) for $(c^n - d^n)$ to be well-defined as $m \to \infty$, it would mean that $m$ in $(c^n - d^n)$ ought to be replaced by a physically equivalent parameter such that $(c^n - d^n)$ continues to be well-defined in this limit (although its dependence on $\beta$ and independence of $\epsilon_{1,2}$ should remain) -- such a parameter is the energy scale $\Lambda$, which is fixed as $m \to \infty$; (iv) (\ref{coherent-A}) and (\ref{psi = delta}) mean that the $n^{\rm th}$ power of  ${\Lambda}$ ought to accompany the state of energy level $n$ in $| G_{U(1)} \rangle$. Altogether, this means that we can also write (\ref{Z5d-G-pure U(1)-initial}) as
\be
\label{Z5d-G-pure U(1)}
\boxed{| G_{U(1)} \rangle =  {\rm exp} \left( -  \sum_{n > 0} {1 \over n} { (\beta \Lambda)^n \over  1 - q^n} \, {a}_{-n} \right) | \emptyset \rangle,  \quad  \langle G_{U(1)} | = \langle \emptyset | {\rm exp} \left(   \sum_{n > 0} {1 \over n} {(\beta \Lambda)^n \over  1 - q^{-n}} \, {a}_{n} \right)}
\ee
where
\be
\boxed{[{a}_p, {a}_n] = p {1- { q}^{|p|} \over 1- { t}^{|p|}} \delta_{p+n, 0}, \quad a_{p > 0} | \emptyset \rangle = 0}
\label{he-pure U(1)}
\ee
and
\be
\label{variables-pure U(1)}
\boxed{q = e^{- i \beta \sqrt{\epsilon_1 \epsilon_2}},  \quad t = e^{- i \beta ( \epsilon_1 + \epsilon_2 + \sqrt{\epsilon_1 \epsilon_2})}}
\ee
Clearly, in arriving at the boxed relations (\ref{Z5d-pure U(1)}), (\ref{Z5d-G-pure U(1)}), (\ref{he-pure U(1)}) and (\ref{variables-pure U(1)}), we have just derived a 5d AGT correspondence for pure $U(1)$!  

Moreover, notice that $| G_{U(1)} \rangle$ takes the same form as~\cite[eqn.~(5.1)]{Awata} (where $(\alpha, \beta)$ in \emph{loc. cit.} correspond to $(\beta \Lambda, 0)$). In other words, $| G_{U(1)} \rangle$ is a \emph{coherent state} in a level one module over a Ding-Iohara algebra. Thus, we have a $U(1)$ version of the $SU(2)$ result in~\cite{Awata-AGT}.

\bigskip\noindent{\it Rederiving the 5d AGT Correspondence for Pure $U(1)$ in the Topological String Limit}

In the topological string limit of $\epsilon_1 =  - \epsilon_2 = \hbar$, we have from (\ref{variables-pure U(1)}), the relation $q=t$. As such, we can write the 5d Nekrasov instanton partition function for pure $U(1)$ in this limit as
\be
\label{Z5d-pure U(1)-h}
 \boxed{Z^{{\rm pure}, \, 5d, \, \hbar}_{{\rm inst}, \, U(1)} (\hbar, \beta, \Lambda) = \langle G^\hbar_{U(1)}| G^\hbar_{U(1)} \rangle}
\ee
where 
\be
\label{Z5d-pure U(1)-Gaiotto-h}
\boxed{\langle G^\hbar_{U(1)}| = \langle 0 | \, {\rm exp} \left(  \sum_{n > 0} {1 \over n} {(\beta \Lambda)^n \over  1 - q^{-n}} \, {\cal J}_{n} \right), \quad  | G^\hbar_{U(1)} \rangle = {\rm exp} \left( - \sum_{n > 0} {1 \over n} { (\beta \Lambda)^n  \over  1 - q^n} \, {\cal J}_{-n} \right) | 0 \rangle, \quad q = e^{ \beta \hbar}}
\ee
and\footnote{In the following, we have ${\cal J}_0 | 0 \rangle = 0$ because the Heisenberg algebra is realized by a single complex chiral fermion on the I-brane ${\cal C}_{\rm eff} = {\bf S}^2$.}
\be
\boxed{[{\cal J}_p, {\cal J}_n] = p \,  \delta_{p+n, 0}, \quad {\cal J}_p | 0 \rangle = 0 \quad {\rm for} \quad p \geq 0}
\label{he-pure U(1)-h}
\ee
Clearly, in arriving at the boxed relations (\ref{Z5d-pure U(1)-h}), (\ref{Z5d-pure U(1)-Gaiotto-h}) and (\ref{he-pure U(1)-h}), we have just rederived a 5d AGT correspondence for pure $U(1)$ in the topological string limit. The above three relations coincide with (\ref{Z5d-tilde})--(\ref{HA-5}), as expected.

\bigskip\noindent{\it  Level One Representation of the Ding-Iohara Algebra and the Double Loop Algebra of $U(1)$}

Just like $ | G_{U(1)} \rangle$,  $ | G^\hbar_{U(1)} \rangle$ is a coherent state or Whittaker vector in a level one module over a Ding-Iohara algebra with $c, d \to 0$, albeit at $q = t$.  Comparing (\ref{Z5d-pure U(1)-h}) with (\ref{Z5d}), we find a correspondence between $ | G^\hbar_{U(1)} \rangle$ and $| {\rm cir'} \rangle$; in turn, (\ref{state double loop}) means that\emph{ a Whittaker vector in a level one module over a Ding-Iohara algebra with $c, d \to 0$ and $q = t$, is also a Whittaker vector in a module over a universal central extension of the double loop algebra of $U(1)$}!

\bigskip\noindent{\it A 5d AGT Correspondence for Pure $SU(N)$}

Likewise, the pure $SU(N)$ case with Coulomb moduli $\vec a = (a_1, \dots, a_N)$ can also be obtained by making the replacements (\ref{Z5d-SU(N)-matter-vacuum})--(\ref{Z5d-SU(N)-matter-vertex operator}) in (\ref{1-point-U(1)}):
\be
\label{1-point to pure}
{Z^{{\rm pure}, \, 5d}_{{\rm inst}, \, SU(N)} (\epsilon_1, \epsilon_2, \vec a, \beta, \Lambda)  =   \langle {\bf 0} |  {\Phi}^{\otimes N}_{m \to \infty} (1) | {\bf 0} \rangle_{{\bf S}^2}}.
\ee 
Recall from the previous subsection that ${\Phi}_{m \to \infty}$ is also a vertex operator of a level one representation of the Ding-Iohara algebra, so ${\Phi}^{\otimes N}_{m \to \infty}$ is a vertex operator of a level $N$ representation of the Ding-Iohara algebra, or rather
\be
\label{q-W-pure-SU(N)}
{{\Phi}^{\otimes N}_{m \to \infty}:  \, \widehat {{\cal W}^q_N} \to \widehat {{\cal W}^q_N}}.
\ee

Explicitly,  
\be
\hspace{-0.3cm}{\Phi}^{\otimes N}_{m \to \infty} (1) = e^{\sum_{n_1 > 0} \left({c^{-n_1} -  d^{-n_1} \over  1 - q^{-n_1}} \, {{a}_{n_1}  \over n_1} -  {c^{n_1} - (t / q)^{n_1} d^{n_1} \over  1 - q^{n_1}} \, {{a}_{-n_1}  \over n_1}\right)} \cdots  e^{\sum_{n_N > 0}  \left({c^{-n_N} -  d^{-n_N} \over  1 - q^{-n_N}} \, {{a}_{n_N}  \over n_N} -  {c^{n_N} - (t / q)^{n_N} d^{n_N} \over  1 - q^{n_N}} \, {{a}_{-n_N} \over n_N}\right)},
\ee
so we can write
\be
\label{Z5d-pure-SU(N)-initial}
Z^{{\rm pure}, \, 5d}_{{\rm inst}, \, SU(N)} (\epsilon_1, \epsilon_2, \vec a, \beta, \Lambda)  = \langle G_{SU(N)} | {\mathscr G}_{SU(N)} | G_{SU(N)} \rangle, 
\ee
where
\be
\label{curly G}
{\mathscr G}_{SU(N)} = e^{\sum_{n_1 >0} {1 \over n_1} {d^{n_1}(t^{n_1} - q^{n_1}) \over q^{n_1}(1 - q^{n_1})} {a}_{-n_1}} \cdots e^{\sum_{n_N >0} {1 \over n_N} {d^{n_N}(t^{n_N} - q^{n_N}) \over q^{n_N}(1 - q^{n_N})} {a}_{-n_N}}, 
\ee
\be
\label{Z5d-pure-SU(N)-G-initial}
\hspace{-0.05cm}{| G_{SU(N)} \rangle =  (e^{-  \sum_{n_1 > 0} {1 \over n_1} {c^{n_1} -  d^{n_1} \over  1 - q^{n_1}} \, {a}_{-{n_1}}} \cdots e^{-  \sum_{n_N > 0} {1 \over n_N} {c^{n_N} -  d^{n_N} \over  1 - q^{n_N}} \, {a}_{-n_N}} ) \cdot \left( | \emptyset \rangle_1 \otimes \cdots \otimes | \emptyset \rangle_N \right)},
\ee
\be
\label{Z5d-pure-SU(N)-G-initial-dual}
\hspace{-0.03cm}{\langle G_{SU(N)} | = \left({_N\hspace{-0.02cm}\langle \emptyset |}\otimes \cdots \otimes{_1\hspace{-0.02cm}\langle \emptyset |} \right) \cdot (e^{\sum_{n_N > 0} {1 \over n_N} {c^{-n_N} -  d^{-n_N} \over  1 - q^{-n_N}} \, {a}_{n_N}} \cdots e^{\sum_{n_1 > 0} {1 \over n_1} {c^{-n_1} -  d^{-n_1} \over  1 - q^{-n_1}} \, {a}_{n_1}})}.
\ee
Here, the $a_{-n_k}$'s and $a_{n_k}$'s act nontrivially only on $| \emptyset \rangle_k$ and ${_k\hspace{-0.02cm}\langle \emptyset |}$, respectively; the rescaling-invariant parameters $c = e^{- \beta ({m \to \infty})}$,  $d = e^{-2\beta  ({m \to \infty})}$, 
${q = e^{- i \beta \sqrt{\epsilon_1 \epsilon_2}}, \quad t = e^{- i \beta ( \epsilon_1 + \epsilon_2 + \sqrt{\epsilon_1 \epsilon_2})}}$;
while  $a_{m_k > 0} | \emptyset \rangle_k = 0$, $_k\hspace{-0.02cm}\langle \emptyset |a_{m_k < 0} = 0$,
and ${{[{a}_{m_k}, {a}_{n_k}] = m_k {1- { q}^{|m_k|} \over 1- { t}^{|m_k|}} \delta_{m_k + n_k, 0}}}$.

Nonetheless, $d \to 0$ so $\mathscr G \to 1$ whence we can write (\ref{Z5d-pure-SU(N)-initial}) as
\be
\label{Z5d-pure-SU(N)}
\boxed{Z^{{\rm pure}, \, 5d}_{{\rm inst}, \, SU(N)} (\epsilon_1, \epsilon_2, \vec a, \beta, \Lambda)  = \langle G_{SU(N)}  | G_{SU(N)} \rangle}
\ee
Also, note that (i) for $n_k > 0$, we have $(c^{-n_k} - c^{n_k}) = (d^{-n_k} -d^{n_k}) \to \infty$, i.e., $(c^{-n_k} - d^{-n_k}) = (c^{n_k} - d^{n_k})$; (ii) since $c$ and $d$ are rescaling-invariant, so must $(c^{n_k} - d^{n_k})$; (iii) for $(c^{n_k} - d^{n_k})$ to be well-defined as $m \to \infty$, it would mean that $m$ in $(c^{n_k} - d^{n_k})$ ought to be replaced by a physically equivalent parameter such that $(c^{n_k} - d^{n_k})$ continues to be well-defined in this limit (although its dependence on $\beta$ and independence of $\epsilon_{1,2}$ should remain) --  such a parameter is the energy scale $\Lambda$, which is fixed as $m \to \infty$; (iv) (\ref{coherent-A}) and (\ref{psi = delta}) mean that the $n_k^{\rm th}$ power of ${\Lambda}$ ought to accompany the state of energy level $n_k$ in $| G_{SU(N)} \rangle$ that is created from $| \emptyset \rangle_k$. Altogether, this means that we can also write (\ref{Z5d-pure-SU(N)-G-initial}) and (\ref{Z5d-pure-SU(N)-G-initial-dual}) as
\be
\label{Z5d-pure-SU(N)-G}
\boxed{\hspace{-0.00cm}{| G_{SU(N)} \rangle =  (e^{-  \sum_{n_1 > 0} {1 \over n_1} { (\beta \Lambda)^{n_1} \over  1 - q^{n_1}} \, {a}_{-{n_1}}} \cdots e^{-  \sum_{n_N > 0} {1 \over n_N} {(\beta \Lambda)^{n_N} \over  1 - q^{n_N}} \, {a}_{-n_N}}) \cdot \, \, \left( | \emptyset \rangle_1 \otimes \cdots \otimes | \emptyset \rangle_N \right)}}
\ee
and
\be
\label{Z5d-pure-SU(N)-G-dual}
\boxed{\hspace{-0.02cm}{\langle G_{SU(N)} | = \left({_N\hspace{-0.02cm}\langle \emptyset |}\otimes \cdots \otimes{_1\hspace{-0.02cm}\langle \emptyset |} \right) \cdot (e^{\sum_{n_N > 0} {1 \over n_N} {(\beta \Lambda)^{n_N} \over  1 - q^{-n_N}} \, {a}_{n_N}} \cdots e^{\sum_{n_1 > 0} {1 \over n_1} {(\beta \Lambda)^{n_1} \over  1 - q^{-n_1}} \, {a}_{n_1}})}}
\ee
where
\be
\boxed{{{[{a}_{m_k}, {a}_{n_k}] = m_k {1- { q}^{|m_k|} \over 1- { t}^{|m_k|}} \delta_{m_k + n_k, 0}}}, \quad a_{m_k > 0} | \emptyset \rangle_k = 0}
\label{he-pure-SU(N)}
\ee
and
\be
\label{variables-pure-SU(N)}
\boxed{q = e^{- i \beta \sqrt{\epsilon_1 \epsilon_2}},  \quad t = e^{- i \beta ( \epsilon_1 + \epsilon_2 + \sqrt{\epsilon_1 \epsilon_2})}}
\ee
Clearly, in arriving at the boxed relations (\ref{Z5d-pure-SU(N)}), (\ref{Z5d-pure-SU(N)-G}), (\ref{Z5d-pure-SU(N)-G-dual}), (\ref{he-pure-SU(N)}) and (\ref{variables-pure-SU(N)}), we have just derived a 5d AGT correspondence for pure $SU(N)$! 

Moreover, (\ref{q-W-pure-SU(N)}), and the fact that $| G_{SU(N)} \rangle $ is a sum over states of all possible energy levels (see (\ref{Z5d-pure-SU(N)-G})), mean that $| G_{SU(N)} \rangle $ is a \emph{coherent state} in  a Verma module over ${\cal W}^q_N$, a $q$-deformed affine ${\cal W}_N$-algebra. Thus, we have an $SU(N)$ generalization of the result for $SU(2)$ in~\cite{Awata-AGT}!

\bigskip\noindent{\it An Equivariant Index of the Dirac Operator on Instanton Moduli Space and a Whittaker Function on a $q$-Deformed Affine ${\cal W}_N$-Algebra}

Recall at this point from $\S$2.1 that since the BPS spectrum ${\cal H}^\Omega_{\rm BPS}$ in (\ref{BPS-AGT-AB})  arises from the \emph{topological} sector of a (non-dynamically) $\bf g$-gauged supersymmetric sigma model on $\Sigma_t = {\bf S}^1 \times \mathbb I_t$, it would coincide with the spectrum of the resulting (non-dynamically) $\bf g$-gauged supersymmetric quantum mechanical  instanton as $\Sigma_t \to {\rm point}$.\footnote{Here, ${\bf g} \in U(1) \times U(1) \times T$, where $T \subset SU(N)$ is a Cartan subgroup.}  As such, $Z'_{{\rm BPS}, m} (\epsilon_1, \epsilon_2, \vec a, \beta \to 0)$ in (\ref{rhs of Z_inst-5d}) can also be regarded as a partition function associated with a (non-dynamically) $\bf g$-gauged supersymmetric quantum mechanical instanton with target  ${\cal U}({\cal M}_{SU(N), m})$. In turn, this means that $Z_{\rm BPS}$ in (\ref{5d BPS}) which underlies $Z^{{\rm pure}, \, 5d}_{{\rm inst}, \, SU(N)}$ in (\ref{Z5d-pure-SU(N)}), can be regarded as a partition function associated with a (non-dynamically) $\bf g$-gauged supersymmetric quantum mechanical instanton  with target  $L{\cal U}({\cal M}_{SU(N)})$, the loop space of ${\cal U}({\cal M}_{SU(N)})$. Consequently, according to~\cite{Ba}, $Z^{{\rm pure}, \, 5d}_{{\rm inst}, \, SU(N)}$ ought to be given by  $Z_{\rm inst}$ in (\ref{Z_inst-5d}) but with the underlying integrand multiplied by $\widehat A^{\bf g}_\beta({\cal U}({\cal M}_{SU(N)}))$, the $\bf g$-equivariant A-roof genus of ${\cal U}({\cal M}_{SU(N)})$ with expansion parameter $\beta$. Hence, we can also write (\ref{Z5d-pure-SU(N)}) as (c.f.~\cite{NN})
\be
 \boxed{\sum_{m = 0}^\infty {\Lambda}^{2mN} \, \oint_{{\cal U}({\cal M}_{SU(N), m})} \widehat A^{{\bf g}(\epsilon_i, \vec a)}_\beta ({\cal U}({\cal M}_{SU(N), m})) = \mathscr W_{G_{SU(N)}}}
\ee
where $\oint_{\cal M} \widehat A^{\bf g}_\beta ({\cal M})$ is the ${\bf g}$-equivariant index of the Dirac operator on $\cal M$, and  $\mathscr W_{G_{SU(N)}}$  is a Whittaker function on ${\cal W}^q_N$.

\bigskip\noindent{\it Rederiving the 5d AGT Correspondence for Pure $SU(N)$ in the Topological String Limit}

In the topological string limit of $\epsilon_1 =  - \epsilon_2 = \hbar$, Omega-deformation on the RHS of (\ref{AGT-M-duality-AB}) effectively vanishes. As explained in $\S$2.2, the partially gauged chiral CFT behind (\ref{AGT-AB-chiral CFT}) would be ungauged, whence the $N$ complex chiral fermions on the I-brane would (i) realize a level one affine algebra $\frak {su}(N)_{\rm aff, 1}$; (ii) generate, via their convenient packing into a single complex chiral fermion $\Psi$ and its bosonization, a Heisenberg algebra.

 From (\ref{variables-pure-SU(N)}), we find that if $\epsilon_1 =  - \epsilon_2 = \hbar$, then $q=t$. In turn, (\ref{he-pure-SU(N)}) becomes ${{[{a}_{m_k}, {a}_{n_k}] = m_k  \delta_{m_k + n_k, 0}}}$, which is also a Heisenberg algebra. Therefore, point (ii) above would mean that we can regard the $N$-tensor product of vacuum states $ | \emptyset \rangle_1 \otimes \cdots \otimes | \emptyset \rangle_N$ as a \emph{single} vacuum state $| p \rangle$, whence there is no longer a distinction between the individual vacuum states $| \emptyset \rangle_k$ and the corresponding operators $a_{n_k}$ that act nontrivially on them. As such, we can write (\ref{curly G})--(\ref{Z5d-pure-SU(N)-G-initial-dual}) in this topological string limit as
\be
\label{curly G-h}
\hspace{-0.4cm}{\mathscr G}^\hbar_{SU(N)} = e^{{ \sum_{n >0} {N \over n} {d^{n}(t^{n} - q^{n}) \over q^{n}(1 - q^{n})} {a}_{-n}}}, \, \,  | G^\hbar_{SU(N)} \rangle =  e^{-  \sum_{n > 0} {1 \over n} {N(c^{n} -  d^{n}) \over  1 - q^{n}} \, a_{-n}} \, | p \rangle,
\, \,  \langle G^\hbar_{SU(N)} | = \langle p | \, e^{\sum_{n > 0} {1 \over n} {N(c^{-n} -  d^{-n}) \over  1 - q^{-n}} \, a_n}.
\ee
Here,  $a_{n > 0} | p \rangle = 0$, $\langle p |a_{n < 0} = 0$, and ${{[{a}_{m}, {a}_{n}] = m \delta_{m + n, 0}}}$.

As before, $d \to 0$ so $\mathscr G \to 1$. Also, note that (i) for $n > 0$, we have $N(c^{-n} - c^n) = N(d^{-n} -d^n) \to \infty$, i.e., $N(c^{-n} - d^{-n}) = N(c^n - d^n)$; (ii) since $c$ and $d$ are rescaling-invariant, so must $N(c^n - d^n)$; (iii) for $N(c^n - d^n)$ to be well-defined as $m \to \infty$, it would mean that $m$ in $N(c^n - d^n)$ ought to be replaced by a physically equivalent parameter such that $N(c^n - d^n)$ continues to be well-defined in this limit (although its dependence on $N, \beta$ and independence of $\epsilon_{1,2}$ should remain)  -- such a parameter is the energy scale $\Lambda$, which is fixed as $m \to \infty$; (iv) (\ref{coherent-A}) and (\ref{psi = delta}) mean that the $n^{\rm th}$ power of  ${\Lambda}^N$ ought to accompany the state of energy level $n$ in $| G^\hbar_{SU(N)} \rangle$. Altogether, this means that we can also write (\ref{Z5d-pure-SU(N)-initial}) in this topological string limit as
\be
\label{Z5d-pure SU(N)-h}
 \boxed{Z^{{\rm pure}, \, 5d, \, \hbar}_{{\rm inst}, \, SU(N)} (\hbar, \vec a, \beta, \Lambda') = \langle G^\hbar_{SU(N)}| G^\hbar_{SU(N)} \rangle}
\ee
where 
\be
\label{Z5d-pure SU(N)-Gaiotto-h}
\hspace{-0.2cm}\boxed{\langle G^\hbar_{SU(N)}| = \langle p | \, {\rm exp} \left(  \sum_{n > 0} {1 \over n} {(\beta \Lambda')^n \over  1 - q^{-n}} \, {\cal J}_{n} \right), \quad  | G^\hbar_{SU(N)} \rangle = {\rm exp} \left( - \sum_{n > 0} {1 \over n} { (\beta \Lambda')^n  \over  1 - q^n} \, {\cal J}_{-n} \right) | p \rangle, \quad q = e^{ \beta \hbar}}
\ee
$\Lambda' = \Lambda^N$, and\footnote{In the following, we have ${\cal J}_0 | p \rangle = p$, where $p \sim \sqrt{{\vec a}^2 / \hbar^2}$ is the energy level of $| p \rangle$, because the Heisenberg algebra is realized by $N$ complex chiral fermions on the I-brane ${\cal C}_{\rm eff} = {\bf S}^2$.}
\be
\boxed{[{\cal J}_m, {\cal J}_n] = m \,  \delta_{m+n, 0}, \qquad {\cal J}_0 | p \rangle = p \sim \sqrt{ \vec a^2 / \hbar^2}, \qquad {\cal J}_m | p \rangle = 0 \quad {\rm for} \quad m > 0}
\label{he-pure SU(N)-h}
\ee
Clearly, in arriving at the boxed relations (\ref{Z5d-pure SU(N)-h}), (\ref{Z5d-pure SU(N)-Gaiotto-h}) and (\ref{he-pure SU(N)-h}), we have just rederived a 5d AGT correspondence for pure $SU(N)$ in the topological string limit. The above three relations coincide with (\ref{Z5d-tilde})--(\ref{HA-5}), as expected.

\bigskip\noindent{\it  Level $N$ Representation of the Ding-Iohara Algebra and the Double Loop Algebra of $SU(N)$}

Just like $ | G_{SU(N)} \rangle$,  $ | G^\hbar_{SU(N)} \rangle$ is a coherent state or Whittaker vector in a level $N$ module over a Ding-Iohara algebra with $c, d \to 0$, albeit at $q = t$.  Comparing (\ref{Z5d-pure SU(N)-h}) with (\ref{Z5d}), we find a correspondence between $ | G^\hbar_{SU(N)} \rangle$ and $| {\rm cir'} \rangle$; in turn, (\ref{state double loop}) means that\emph{ a Whittaker vector in a level $N$ module over a Ding-Iohara algebra with $c, d \to 0$ and $q = t$, is also a Whittaker vector in a module over a universal central extension of the double loop algebra of $SU(N)$}!

\newsection{A 5d ``Fully-Ramified'' AGT Correspondence and Relativistic Integrable Systems}

\newsubsection{An M-Theoretic Derivation of a 4d ``Fully-Ramified''  Pure AGT Correspondence: A Review}

Let us now review the M-theoretic derivation of a 4d ``fully-ramified'' AGT correspondence for pure $SU(N)$ in~\cite{4d AGT}. 

To this end, note that the ``fully-ramified'' version of the AGT correspondence for pure $SU(N)$ can be obtained by considering the following duality setup which is a slight variant of (\ref{AGT-M-duality-AB}):
\be
\underbrace{\mathbb R^4\vert_{\epsilon_1, \epsilon_2}  \times \Sigma_{t}}_{\textrm{$N$ M5-branes + 4d full defect}}\times \mathbb R^{5}\vert_{\epsilon_3; \,  x_{6,7}}  \quad \Longleftrightarrow  \quad   {\mathbb R^{5}}\vert_{\epsilon_3; \, x_{4,5}} \times \underbrace{{\cal C}  \times TN_N^{R\to 0}\vert_{\epsilon_3; \, x_{6,7}}}_{\textrm{$1$ M5-branes + 4d full defect}}.
\label{AGT-M-duality-AB-defect}
\ee
Here, the worldvolume 4d full defects are effected by an appropriate orbifold background (see~\cite[$\S$2.3]{4d AGT}); we have a common half-BPS boundary condition at the tips of $\mathbb I_t \subset \Sigma_{t} = {\bf S}^1 \times \mathbb I_t$; the radius of ${\bf S}^1$ is $\beta$; $\mathbb I_t \ll \beta$; $\cal C$ is \emph{a priori} the same as $\Sigma_{t} $; the 4d worldvolume full defect on the LHS of (\ref{AGT-M-duality-AB-defect}) wraps $\Sigma_{t}$ and the $z$-plane in $\mathbb R^4\vert_{\epsilon_1, \epsilon_2} \simeq \mathbb C_z\vert_{\epsilon_1} \times \mathbb C_w\vert_{\epsilon_2}$; the\emph{ dual} 4d worldvolume full defect on the RHS of (\ref{AGT-M-duality-AB-defect})  wraps $\cal C$ and the $x_8$-$x_9$ directions in $TN_N^{R\to 0}\vert_{\epsilon_3; \, x_{6,7}}$, where the $x_9$-direction is spanned by the ${\bf S}^1$-fiber of  $TN_N^{R\to 0}\vert_{\epsilon_3; \, x_{6,7}}$, while the $x_6$-$x_7$-$x_8$-directions are spanned by its $\mathbb R^3\vert_{\epsilon_3; \, x_{6,7}}$ base; and the $\epsilon_i$'s are parameters of the Omega-deformation along the indicated planes described in (\ref{table-AGT}).

\bigskip\noindent{\it The Spectrum of Spacetime BPS States on the LHS of (\ref{AGT-M-duality-AB-defect})}

Let us now determine the spectrum of spacetime BPS states on the LHS of (\ref{AGT-M-duality-AB-defect})  that defines a ``fully-ramified'' generalization of the partition function in (\ref{5d BPS}). As explained in detail in \cite[$\S$6.1]{4d AGT}, we can express the Hilbert space ${\cal H}^{\Omega}_{\rm BPS}$ of spacetime BPS states on the LHS of  (\ref{AGT-M-duality-AB-defect}) as
\be
{\cal H}^\Omega_{\rm BPS} = \bigoplus_{a'} {\cal H}^\Omega_{{\rm BPS}, a'}  =  \bigoplus_{a'} ~{\rm IH}^\ast_{U(1)^2 \times T} \, {\cal U}({\cal M}_{SU(N), T, a'}), 
\label{BPS-AGT-AB-defect}
\ee
where ${\rm IH}^\ast_{U(1)^2 \times T} \, {\cal U}({\cal M}_{SU(N), T, a'})$ is the  $U(1)^2 \times T$-equivariant intersection cohomology of the Uhlenbeck compactification ${\cal U}({\cal M}_{SU(N), T, a'})$ of the (singular) moduli space ${\cal M}_{SU(N), T, a'}$ of ``fully-ramified'' $SU(N)$-instantons on $\mathbb R^4$ with ``fully-ramified'' instanton number $a'$. Here, the positive number $a' = a + {\rm Tr} \, \alpha\frak m$, where $a$ is the ordinary instanton number; Tr is a quadratic form on ${\frak {su}}(N)$; $\alpha \in \frak t$ is the holonomy parameter that is the commutant of the Cartan subgroup $T \subset SU(N)$; $\frak t$ is the Lie algebra of  $T$; ${\frak m} \in \Lambda_{\rm cochar}$ is the ``magnetic charge''; and $ \Lambda_{\rm cochar}$ is the cocharacter lattice of $SU(N)$.

\bigskip\noindent{\it The Spectrum of Spacetime BPS States on the RHS of (\ref{AGT-M-duality-AB-defect})}

Let us next ascertain the corresponding spectrum of spacetime BPS states on the RHS of (\ref{AGT-M-duality-AB-defect}). According to~\cite[$\S$6.1]{4d AGT}, the spacetime BPS states would be furnished by the ``fully-ramified'' I-brane theory in the following type IIA configuration:
\be
\qquad \textrm{IIA}: \quad \underbrace{ {\mathbb R}^5\vert_{\epsilon_3; x_{4,5}} \times {\cal C} \times {\mathbb R}^3\vert_{\epsilon_3; x_{6,7}}}_{\textrm{I-brane on ${\cal C} = N \textrm{D6} \cap 1\textrm{D4}\cap{\rm {3d \, full \, defect}}$}}.
\label{equivalent IIA system 1 - AGT-AB-defect}
\ee
Here, we have a stack of $N$ coincident D6-branes whose worldvolume is given by ${\mathbb R}^5\vert_{\epsilon_3; x_{4,5}} \times {\cal C}$; a single D4-brane whose worldvolume is given by ${\cal C} \times \mathbb R^3\vert_{\epsilon_3; x_{6,7}}$; and a 3d worldvolume full defect which wraps ${\cal C}$ and the $x_8$-direction in $\mathbb R^3\vert_{\epsilon_3; x_{6,7}} = \mathbb R \times \mathbb R^2\vert_{\epsilon_3}$. 

As explained in detail in~\cite[$\S$6.1]{4d AGT}, from (\ref{equivalent IIA system 1 - AGT-AB-defect}), we find that in place of (\ref{AGT-AB-H=W}), the Hilbert space ${\cal H}^{\Omega'}_{\rm BPS}$ of space-time BPS states on the RHS of (\ref{AGT-M-duality-AB-defect}) would be given by 
\be
\label{coset-AGT-A-full}
{\cal H}^{\Omega'}_{\rm BPS} = \widehat {{\frak {su}(N)}}_{{\rm aff}, k},
\ee
where $\widehat {{\frak {su}(N)}}_{{\rm aff}, k}$ is an integrable module over the affine Lie algebra ${{\frak {su}(N)}}_{{\rm aff}, k}$ of level $k$. 

\bigskip\noindent{\it A `` Fully-Ramified'' AGT Correspondence for Pure $SU(N)$}

Clearly, the physical duality of the compactifications in (\ref{AGT-M-duality-AB-defect}) will mean that ${\cal H}^\Omega_{\rm BPS}$ in (\ref{BPS-AGT-AB-defect}) is equivalent to ${\cal H}^{\Omega'}_{\rm BPS}$ in (\ref{coset-AGT-A-full}), i.e.,
\be
\bigoplus_{a'} ~{\rm IH}^\ast_{U(1)^2 \times T} \, {\cal U}({\cal M}_{SU(N), T, a'}) = \widehat {{\frak {su}(N)}}_{{\rm aff}, k},
 \label{AGT-duality-A-defect}
\ee
where the level $k$ and central charge $c$ of $\widehat {{\frak {su}(N)}}_{{\rm aff}, k}$ are given by
\be
\label{K-ab}
{k = - N  - {\epsilon_2 \over \epsilon_1}}, \qquad 
{ c = {\epsilon_1 \over \epsilon_2} (N^3 - N) + N^2 -1}. 
\ee

Consequently, in place of (\ref{q | q}), we have 
\be
\label{q | q, full-defect-AB}
{Z_{\rm inst} (\Lambda, \epsilon_1, \epsilon_2, \vec a, T) = \langle {\rm coh}_T | {\rm coh}_T \rangle},
\ee
where
\be
\label{coherent-AB-full-defect}
{ |{\rm coh}_T \rangle = \bigoplus_{a'}  {\cal G}^{a'}  | \Psi_{a', T} \rangle}.
\ee
Here, $\vec a$ is the Coulomb moduli of the underlying pure $SU(N)$ theory along $\mathbb R^4\vert_{\epsilon_1, \epsilon_2 }$ on the LHS of (\ref{AGT-M-duality-AB-defect}); $ |{\rm coh}_T \rangle \in \widehat {{\frak {su}(N)}}_{{\rm aff}, k}$; ${\cal G}^{a'}$ is some real constant depending on the energy scale $\Lambda$; and $ | \Psi_{a', T} \rangle \in {\rm IH}^\ast_{U(1)^2 \times T}  {\cal U}({\cal M}_{SU(N), T, a'})$ is also a state in $\widehat {{\frak {su}(N)}}_{{\rm aff}, k}$ with energy level $n_{a'}$ determined by the ``fully-ramified'' instanton number $a'$. 

Like $ |{\rm coh} \rangle$ and $\langle {\rm coh}|$,  $ |{\rm coh}_T \rangle$ and $\langle {\rm coh}_T |$ ought to be a state and its dual associated with the puncture at $z = 0$ and $z=\infty$ in ${\cal C} = {\bf S}^2$, respectively. Furthermore, as the RHS of (\ref{coherent-AB-full-defect}) is a sum over states of all possible energy levels, it must mean that $ |{\rm coh}_T \rangle$ is actually a \emph{coherent state}. 

In arriving at the relations (\ref{AGT-duality-A-defect}),  (\ref{K-ab}),  (\ref{q | q, full-defect-AB}) and (\ref{coherent-AB-full-defect}), we have just derived a  ``fully-ramified'' AGT correspondence for pure $SU(N)$. 

\newsubsection{A 5d ``Fully-Ramified'' Pure AGT Correspondence and a Relativistic Periodic Toda Integrable System}

 \bigskip\noindent{\it The ``Fully-Ramified'' 4d Pure Nekrasov Instanton Partition Function in the NS Limit and a Quantum Affine Toda System}

Let us consider the NS limit $\epsilon_2 = 0$ in our derivation of the 4d ``fully-ramified'' AGT correspondence for pure $SU(N)$ in the last subsection. From (\ref{K-ab}), we find that in place of (\ref{q | q, full-defect-AB}), the ``fully-ramified'' 4d pure Nekrasov instanton partition function is now
\be
\label{q | q, full defect-NS limit}
{Z_{\rm inst} (\Lambda, \epsilon_1, 0, \vec a, T) = \langle {\rm coh}^{\rm crit}_T | {\rm coh}^{\rm crit}_T \rangle},
\ee
where $\vec a$ is the Coulomb moduli of the underlying pure $SU(N)$ theory along $\mathbb R^4\vert_{\epsilon_1, \epsilon_2 = 0}$ on the LHS of (\ref{AGT-M-duality-AB-defect}); $T \subset SU(N)$ is the maximal torus that $SU(N)$ reduces to along $\mathbb R^2\vert_{\epsilon_1} \subset \mathbb R^4\vert_{\epsilon_1, \epsilon_2 = 0}$; the coherent state $| {\rm coh}^{\rm crit}_T \rangle \in {\widehat {\frak {su}(N)}}_{{\rm aff}, {\rm crit}}$; and ${\widehat {\frak {su}(N)}}_{{\rm aff}, {\rm crit}}$ is an integrable module over the affine Lie algebra ${\frak {su}(N)}_{{\rm aff}, {\rm crit}}$ at the \emph{critical level} $k = -N$.

Via the dimension-one currents $J_{a_i}$ that generate ${\frak {su}(N)}_{{\rm aff}, {\rm crit}}$ on ${\cal C} = {\bf S}^2$, one can define the Segal-Sugawara operators  
\be 
S^{(s_i)}(z) = (k+h^{\vee}) T^{(s_i)}(z), \quad s_i = i +1, \quad i = 1, 2, \dots, N-1,
\label{act 2}
\ee 
where $h^\vee$ is the dual Coxeter number of $\frak {su}(N)$, the spin-$s_i$ operators $T^{(s_i)}$ are just higher spin generalizations of the holomorphic stress tensor $T^{(2)}$, and
\be S^{(s_i)}(z) = :{
d}^{a_1 a_2 a_3 \dots a_{s_i}}(k) (J_{a_1} J_{a_2}\dots
J_{a_{s_i}})  (z):,
\label{S^{(s_i)}(z)} 
\ee 
where ${d}^{a_1 a_2 a_3 \dots a_{s_i}}(k)$ is a completely symmetric traceless $\frak {su}(N)$-invariant tensor of rank $s_i$ (which depends on the level $k$); in other words, the $N-1$ number of $S^{(s_i)}$'s are Casimir operators.

 From (\ref{act 2}), one can see that the $S^{(s_i)}$'s generate in their OPE's with all other operators of
the quantum CFT on $\cal C$,  $(k+h^{\vee})$ times the field transformations generated by the $T^{(s_i)}$'s. Therefore, at the critical level $k=-N = -h^\vee$, the $S^{(s_i)}$'s  generate \emph{no} field transformations at all: their OPE's amongst themselves, and with all other field operators, are regular, and they are said to span the central elements of (the universal enveloping algebra of) the affine $SU(N)$-algebra at the critical level. Hence, on any correlation function of operators, the $S^{(s_i)}$'s effectively act as $c$-numbers. In particular, this means that the RHS of (\ref{q | q, full defect-NS limit}) -- which can be interpreted as a two-point correlation function of coherent state operators -- is a simultaneous eigenfunction of the commuting $S^{(s_i)}$'s. As the $S^{(s_i)}$'s generically act as  order-$s_i$ differential operators in their action on a correlation function of primary state operators (see for example~\cite[$\S$15.7]{CFT text}), and since a coherent state can be obtained by applying creation operators on a primary state, i.e., a coherent state operator can be derived from a primary state operator, our discussion hitherto would mean that $Z_{\rm inst} (\Lambda, \epsilon_1, 0, \vec a, T)$ ought to be a simultaneous eigenfunction of $N-1$ commuting differential operators $D_1, D_2, \dots D_{N-1}$ derived from the Casimirs of an affine $SU(N)$-algebra.  

Indeed, note that the coherent state $| {\rm coh}^{\rm crit}_T \rangle$ is also known as a Whittaker vector in representation theory whence its inner product on the RHS of (\ref{q | q, full defect-NS limit}) is a Whittaker function associated with ${\frak {su}(N)}_{{\rm aff}, {\rm crit}}$; in turn, according to~\cite[$\S$2]{etingof}, this would mean that  $Z_{\rm inst} (\Lambda, \epsilon_1, 0, \vec a, T)$ must be a simultaneous eigenfunction  of $N-1$ quantum Toda Hamiltonians ${\cal D}^{(1)}_{\rm Toda}, {\cal D}^{(2)}_{\rm Toda}, \dots, {\cal D}^{(N-1)}_{\rm Toda}$ that are associated with an affine $SU(N)$-algebra, i.e., 
\be
\label{toda IS}
{{\cal D}^{(l)}_{\rm Toda} \cdot Z_{\rm inst} (\Lambda, \epsilon_1, 0, \vec a, T) = {\cal E}^{(l)}_{\rm Toda}  \, Z_{\rm inst} (\Lambda, \epsilon_1, 0, \vec a, T) },
\ee
where the ${\cal D}^{(l)}_{\rm Toda}$'s are commuting Casimir differential operators; the ${\cal E}^{(l)}_{\rm Toda}$'s are complex eigenvalues; and $l = 1, 2, \dots, N-1$.  Furthermore, it has also been shown in~\cite{MW} that the spectral curve of the quantum affine Toda system defined by (\ref{toda IS}), is just $\Sigma_{SW}$, the 4d Seiberg-Witten curve.

Note that this is also consistent with~\cite[Corollary~3.7(2)]{J-function}, where ${\cal Z}^{\rm aff}_{G, B}$ and $\epsilon$ in \emph{loc.~cit.} correspond respectively to $Z_{\rm inst} (G, \epsilon_1, 0, \vec a, T)$  and $\epsilon_2$ here.

\bigskip\noindent{\it The 5d Seiberg-Witten Curve}

Now recall that the I-brane consisting of $N$ D6-branes and a single D4-brane wrapping ${\cal C}$ can be regarded as an $N$-fold cover of ${\cal C}$ which, in the 4d case, can be interpreted as $\Sigma_{SW}$ (see explanations leading up to (\ref{4d SW curve})). What is its corresponding interpretation in the 5d case? 

Similar to $\S$2.2, we have on $\cal C$, a (chiral) $SU(N)$ WZW model (which underlies ${\frak {su}(N)}_{{\rm aff}, {\rm crit}}$). As such, according to our explanations in the paragraph leading up to (\ref{state double loop}),  one can equivalently study the 4d theory with gauge group $LSU(N)$ instead of $SU(N)$. This amounts to replacing~\cite{NN-5d} the 4d Higgs field $\phi$  with a first-order differential operator on a circle: $\phi \to \partial_t + A_t + i\varphi$, where $A_t$ and $\varphi$ are an $SU(N)$ gauge field and a real scalar along the circle with coordinate $t$ (in the adjoint representation of $SU(N)$). Correspondingly, we ought to also make this replacement in $\Sigma_{SW}$ of (\ref{4d SW curve}), where $A_t$ and $\varphi$ are here an $SU(N)$ gauge field and a real scalar along a circle $C_\beta$ in $\cal C$ of radius $\beta$ (in the adjoint representation of $SU(N)$) that originate from the $SU(N)$ gauge theory on the $N$ D6-branes in (\ref{equivalent IIA system 1 - AGT-AB-defect}).  

Note at this point that we can also parameterize $\Sigma_{SW}$ as
\be
\Sigma_{SW}: \, \, y^2 = P(x)^2 - \Lambda^{2N},
\ee
where $x,y \in \mathbb C$, and $P(x) = {\rm det}(\phi - x)$. Thus, in the 5d case, the curve ought to be given by 
\be
\label{above}
\Sigma^{\rm {5d}}_{SW}: \, \, y^2= {\rm det}( \partial_t + A_t + i\varphi - x)^2 - \Lambda^{2N}. 
\ee
The determinant ${\rm det}( \partial_t + A_t + i\varphi - x)$ can be computed via zeta-regularization as 
\be
{\rm det}( \partial_t + A_t + i\varphi - x) = \prod_i {1 \over \beta} \, {\rm sinh} (2\pi \beta a_i - x), 
\ee
where the $a_i$'s are the eigenvalues of $\partial_t + A_t + i\varphi$ when $\beta \to 0$. Substituting this in (\ref{above}), we get, after an appropriate change of variables,
\be
\label{SW-5d}
\boxed{\Sigma^{\rm {5d}}_{SW}:  \, \, z + {\Lambda^{2N} \over z}= \prod_i {1 \over \beta} \, {\rm sinh}(2\pi \beta a_i - x)}
\ee
Using the fact that $\sum_i a_i = 0$, one can, after another change of variables, show that $\Sigma^{\rm {5d}}_{SW}$ is the spectral curve of a relativistic periodic Toda integrable system~\cite{NN-5d}.

\bigskip\noindent{\it The ``Fully-Ramified'' 5d Pure Nekrasov Instanton Partition Function in the NS Limit and a Relativistic Periodic Toda Integrable System}
 
Hence, the 5d interpretation of  $\Sigma_{SW}$ is a spectral curve $\Sigma^{\rm {5d}}_{SW}$  of a relativistic periodic Toda integrable system or equivalently, a $q$-deformed quantum affine Toda system~\cite[$\S$6]{etingof}. In the limit $q \to 1$, a $q$-deformed quantum affine Toda system would reduce to the earlier-mentioned quantum affine Toda system~\cite[$\S$6]{etingof}. In the limit $q \to 1$, it has also been rigorously established in~\cite{etingof-ding} that the $N-1$ commuting difference operators ${\cal D}^{(1)}_{q-{\rm Toda}}, {\cal D}^{(2)}_{q-{\rm Toda}}, \dots, {\cal D}^{(N-1)}_{q-{\rm Toda}}$ which define a $q$-deformed quantum affine Toda system, would reduce to the $N-1$ commuting differential operators ${\cal D}^{(1)}_{\rm Toda}, {\cal D}^{(2)}_{\rm Toda}, \dots, {\cal D}^{(N-1)}_{\rm Toda}$ which define the earlier-mentioned quantum affine Toda system. If we conveniently identify $q = e^\beta$, where $\beta$ is the radius of the fifth dimension circle, then in the limit $q \to 1$, we  also have $\Sigma^{\rm {5d}}_{SW} \to \Sigma_{SW}$ and  $Z^{{\rm pure}, \, 5d}_{{\rm inst}, \, SU(N)} ( \Lambda, \epsilon_1, 0, \vec a, T, \beta) \to Z_{\rm inst} (\Lambda, \epsilon_1, 0, \vec a, T) $, where $Z^{{\rm pure}, \, 5d}_{{\rm inst}, \, SU(N)} ( \Lambda, \epsilon_1, 0, \vec a, T, \beta)$ is the ``fully-ramified'' 5d pure Nekrasov instanton partition function in the  $\epsilon_2 = 0 \neq \epsilon_1$ NS limit. These statements hold vice-versa, so this correspondence can be summarized in the diagram below.

 \tikzset{node distance=6cm, auto}
 \begin{center}
 \begin{tikzpicture}
  \node (P) {$\{Z_{\rm inst}; \, \Sigma_{SW}; \, {\cal D}^{(l)}_{\rm Toda}\}$};
  \node (B) [right of=P] {$\{Z^{{\rm pure}, \, 5d}_{{\rm inst}}; \, \Sigma^{\rm {5d}}_{SW}; \, {\cal D}^{(l)}_{q-{\rm Toda}} \}$};
   \draw[transform canvas={yshift=0.5ex},->] (P) --(B) node[midway] {$q \nrightarrow 1$};
\draw[transform canvas={yshift=-0.5ex},->](B) -- (P) node[midway] {$q \to 1$}; 
 \end{tikzpicture}
 \end{center}
 From the diagram, it is clear that the 5d version of  (\ref{toda IS}) must be
\be
\label{q-toda IS}
\boxed{{{\cal D}^{(l)}_{q-{\rm Toda}} \cdot  Z^{{\rm pure}, \, 5d}_{{\rm inst}, \, SU(N)} ( \Lambda, \epsilon_1, 0, \vec a, T, \beta)= {\cal E}^{(l)}_{q-{\rm Toda}}  \, Z^{{\rm pure}, \, 5d}_{{\rm inst}, \, SU(N)} ( \Lambda, \epsilon_1, 0, \vec a, T, \beta)}}
\ee
where the ${\cal E}^{(l)}_{q-{\rm Toda}}$'s are complex eigenvalues, and $l = 1, 2, \dots, N-1$.  In other words, $Z^{{\rm pure}, \, 5d}_{{\rm inst}, \, SU(N)} ( \Lambda, \epsilon_1, 0, \vec a, T, \beta)$ must be a simultaneous eigenfunction of the $N-1$ commuting difference operators ${\cal D}^{(1)}_{q-{\rm Toda}}, {\cal D}^{(2)}_{q-{\rm Toda}}, \dots, {\cal D}^{(N-1)}_{q-{\rm Toda}}$ which define a relativistic periodic Toda integrable system. 
 
 \bigskip\noindent{\it A 5d ``Fully-Ramified'' AGT Correspondence for Pure $SU(N)$ in the NS Limit}

Note that the ${\cal D}^{(l)}_{q-{\rm Toda}}$'s correspond to the central elements of a quantum affine $SU(N)$-algebra at the critical level, while the ${\cal D}^{(1)}_{\rm Toda}$'s correspond to the central elements of (the universal enveloping algebra of) an affine $SU(N)$-algebra at the critical level. Thus, from (\ref{q-toda IS}), (\ref{toda IS}) and (\ref{q | q, full defect-NS limit}), one can conclude that 
\be
\label{Z5d-full defect}
\boxed{Z^{{\rm pure}, \, 5d}_{{\rm inst}, \, SU(N)} ( \Lambda, \epsilon_1, 0, \vec a, T, \beta) = \langle {\rm coh}^{\rm crit}_{T, \beta} | {\rm coh}^{\rm crit}_{T, \beta} \rangle}
\ee 
where
\be
\label{Z5d-full defect-G}
\boxed{| {\rm coh}^{\rm crit}_{T, \beta} \rangle \in  \widehat{U_q}( {\frak {su}(N)_{\rm aff}})_{\rm crit}, \, \, \,  q = e^\beta}
\ee
and $\widehat{U_q}( {\frak {su}(N)_{\rm aff}})_{\rm crit}$ is the integrable module over ${U_q}( {\frak {su}(N)_{\rm aff}})_{\rm crit}$, a quantum affine $SU(N)$-algebra at the critical level. Like $ |{\rm coh}^{\rm crit}_T \rangle$ and $\langle {\rm coh}^{\rm crit}_T |$ in (\ref{q | q, full defect-NS limit}), $ |{\rm coh}^{\rm crit}_{T, \beta} \rangle$ and $\langle {\rm coh}^{\rm crit}_{T, \beta} |$ can be regarded as a state and its dual associated with the puncture at $z = 0$ and $z=\infty$ in ${\cal C} = {\bf S}^2$, respectively. Furthermore, as $ |{\rm coh}^{\rm crit}_{T, \beta} \rangle$  is a 5d analog of $ |{\rm coh}^{\rm crit}_T \rangle$, i.e., $ |{\rm coh}^{\rm crit}_{T, \beta} \rangle$ contains higher excited states beyond those present in the coherent state $ |{\rm coh}^{\rm crit}_T \rangle$ (see explanations after (\ref{Z5d})), it must mean that $ |{\rm coh}^{\rm crit}_{T, \beta} \rangle$ is also a \emph{coherent state}. This is consistent with the mathematical result in~\cite[$\S$4]{etingof} that the simultaneous eigenfunction of the ${\cal D}^{(l)}_{q-{\rm Toda}}$'s is necessarily a Whittaker function on ${U_q}( {\frak {su}(N)_{\rm aff}})_{\rm crit}$. 

Clearly, in arriving at the boxed relations (\ref{Z5d-full defect}) and (\ref{Z5d-full defect-G}), we have just derived a  5d ``fully-ramified'' AGT correspondence for pure $SU(N)$ in the NS limit!

\newsubsection{A 5d ``Fully-Ramified'' AGT Correspondence with Matter and a Relativistic Elliptic Calogero-Moser System}

 \bigskip\noindent{\it The ``Fully-Ramified'' 4d Nekrasov Instanton Partition Function with Single Adjoint Matter in the NS Limit and an Elliptic Calogero-Moser System}

Like the $U(1)$ case mentioned in the sections before, a pure $SU(N)$ theory can also be interpreted as the $m \to \infty$, $q' = e^{2 \pi i \tau'} \to 0$ limit of an $SU(N)$ theory with an adjoint hypermultiplet matter of mass $m$ and complexified gauge coupling $\tau'$, where $m^N e^{   2\pi i \tau'} = \Lambda^N$ remains fixed.  Moreover, note that in this limit, the Hamiltonian and spectral curve of the elliptic Calogero-Moser system associated with the aforementioned theory with adjoint matter, reduce to the Hamiltonian and spectral curve of the quantum affine Toda system~\cite{Phong}.\footnote{See also~\cite{Ito-Morozov}.} 

The above two points mean that in the reverse limit, we have, in (\ref{toda IS}),  the transformation (i) $Z_{\rm inst} (\Lambda, \epsilon_1, 0, \vec a, T) \to Z^{\rm adj}_{\rm inst} (q', \epsilon_1, 0, \vec a, m, T)$, where $Z^{\rm adj}_{\rm inst} (q', \epsilon_1, 0, \vec a, m, T)$ is the ``fully-ramified'' 4d Nekrasov instanton partition function in the NS  limit $\epsilon_2 = 0$ of the $SU(N)$ theory on $\mathbb R^4\vert_{\epsilon_1, \epsilon_2 = 0}$ with an adjoint hypermultiplet matter of mass $m$, $\vec a$ are its Coulomb moduli, and $T \subset SU(N)$ is the maximal torus that $SU(N)$ reduces to along $\mathbb R^2\vert_{\epsilon_1} \subset \mathbb R^4\vert_{\epsilon_1, \epsilon_2 = 0}$; (ii) ${\cal D}^{(l)}_{\rm Toda} \to {\cal D}^{(l)}_{\rm CM}$, where the ${\cal D}^{(l)}_{\rm CM}$'s are differential operators that correspond to the quantum Hamiltonians of the elliptic Calogero-Moser system. In other words, we can write 
\be
\label{CM IS}
{{\cal D}^{(l)}_{\rm CM} \cdot Z^{\rm adj}_{\rm inst} (q', \epsilon_1, 0, \vec a, m, T) = {\cal E}^{(l)}_{\rm CM}  \, Z^{\rm adj}_{\rm inst} (q', \epsilon_1, 0, \vec a, m, T)},
\ee    
where the ${\cal E}^{(l)}_{\rm CM}$'s are  complex eigenvalues, and  $l = 1, 2, \dots, N-1$. In the case where $N = 2$, the eigenvalue ${\cal E}^{(1)}_{\rm CM}$ has been determined explicitly in~\cite[$\S$4.3]{Alday-Tachikawa}.

 \bigskip\noindent{\it The ``Fully-Ramified'' 5d Nekrasov Instanton Partition Function with Single Adjoint Matter in the NS Limit and a Relativistic Elliptic Calogero-Moser System}

Likewise, in the above reverse limit, (i) $Z^{{\rm pure}, \, 5d}_{{\rm inst}, \, SU(N)} ( \Lambda, \epsilon_1, 0, \vec a, T, \beta) \to  Z^{{\rm adj}, \, 5d}_{{\rm inst}, \, SU(N)} (q', \epsilon_1, 0, \newline \vec a,  m, T, \beta)$, where $Z^{{\rm pure}, \, 5d}_{{\rm inst}, \, SU(N)} ( \Lambda, \epsilon_1, 0, \vec a, T, \beta)$ is the ``fully-ramified'' 5d pure Nekrasov instanton partition function while $Z^{{\rm adj}, \, 5d}_{{\rm inst}, \, SU(N)} (q', \epsilon_1, 0, \vec a, m, T, \beta)$ is the ``fully-ramified'' 5d Nekrasov instanton partition function with single adjoint matter of mass $m$, both in the NS limit; (ii) ${\cal D}^{(l)}_{q-{\rm Toda}} \to {\cal D}^{(l)}_{q-{\rm CM}}$~\cite[$\S$7]{etingof}, where the ${\cal D}^{(l)}_{q-{\rm Toda}}$'s and ${\cal D}^{(l)}_{q-{\rm CM}}$'s are commuting difference operators which define a relativistic periodic Toda and elliptic Calogero-Moser system, respectively.  In the $q = e^\beta \to 1$ limit, where $\beta$ is the radius of the fifth dimension circle, (i) $Z^{{\rm pure}, \, 5d}_{{\rm inst}, \, SU(N)} ( \Lambda, \epsilon_1, 0, \vec a, T, \beta) \to Z_{\rm inst} (\Lambda, \epsilon_1, 0, \vec a, T)$; (ii) $Z^{{\rm adj}, \, 5d}_{{\rm inst}, \, SU(N)} (q', \epsilon_1, 0, \vec a,  m, T, \beta) \to Z_{\rm inst} (q', \epsilon_1, 0, \vec a, m, T) = Z^{\rm adj}_{\rm inst}$; (iii) $\{{\cal D}^{(l)}_{q-{\rm Toda}}, {\cal D}^{(l)}_{q-{\rm CM}}\} \to \{{\cal D}^{(l)}_{{\rm Toda}}, {\cal D}^{(l)}_{{\rm CM}}\}$. These statements hold vice-versa, so this correspondence can be summarized in the diagram below.

 \tikzset{node distance=4.7cm, auto}
\begin{center}
 \begin{tikzpicture}
  \node (P) {$\{Z^{{\rm pure}, \, 5d}_{{\rm inst}}; {\cal D}^{(l)}_{q-{\rm Toda}} \}$};
  \node (B) [right of=P] {$\{Z^{{\rm adj}, \, 5d}_{{\rm inst}}; {\cal D}^{(l)}_{q-{\rm CM}}\}$};
  \node (A) [below of=P] {$\{ Z_{\rm inst};  {\cal D}^{(l)}_{\rm Toda} \}$};
  \node (C) [below of=B] {$\{  Z^{\rm adj}_{\rm inst};  {\cal D}^{(l)}_{\rm CM}  \}$};
   \draw[transform canvas={yshift=0.5ex},->] (P) --(B) node[midway] {\footnotesize{$m \nrightarrow \infty$}};
\draw[transform canvas={yshift=-0.5ex},->](B) -- (P) node[midway] {\footnotesize{$m \to \infty$}}; 
 \draw[transform canvas={xshift=0.5ex},->] (P) --(A) node[midway] {$q \to 1$};
\draw[transform canvas={xshift=-0.5ex},->](A) -- (P) node[midway] {$q \nrightarrow 1$}; 
 \draw[transform canvas={yshift=0.5ex},->] (A) --(C) node[midway] {\footnotesize{$m \nrightarrow \infty$}};
\draw[transform canvas={yshift=-0.5ex},->](C) -- (A) node[midway] {\footnotesize{$m \to \infty$}}; 
\draw[transform canvas={xshift=0.5ex},->] (B) --(C) node[midway] {$q \to 1$};
\draw[transform canvas={xshift=-0.5ex},->](C) -- (B) node[midway] {$q \nrightarrow 1$};   \end{tikzpicture}
  \end{center}
From the diagram, and (\ref{toda IS}), (\ref{q-toda IS}) and (\ref{CM IS}), it is clear that we ought to have
\be
\label{q-CM IS}
\boxed{{\cal D}^{(l)}_{q-{\rm CM}} \cdot Z^{{\rm adj}, \, 5d}_{{\rm inst}, \, SU(N)} (q', \epsilon_1, 0, \vec a,  m, T, \beta)  = {\cal E}^{(l)}_{q-{\rm CM}}  \, Z^{{\rm adj}, \, 5d}_{{\rm inst}, \, SU(N)} (q', \epsilon_1, 0, \vec a,  m, T, \beta) }
\ee   
where the ${\cal E}^{(l)}_{q-{\rm CM}}$'s are complex eigenvalues, and $l = 1, 2, \dots, N-1$. In other words, $Z^{{\rm adj}, \, 5d}_{{\rm inst}, \, SU(N)} (q', \epsilon_1, 0, \vec a,  m, T, \beta)$ must be a simultaneous eigenfunction of the $N-1$ commuting difference operators ${\cal D}^{(1)}_{q-{\rm CM}}, {\cal D}^{(2)}_{q-{\rm CM}}, \dots, {\cal D}^{(N-1)}_{q-{\rm CM}}$ which define a relativistic elliptic Calogero-Moser integrable system.

 \bigskip\noindent{\it A 5d ``Fully-Ramified'' AGT Correspondence for $SU(N)$ with Single Adjoint Matter in the NS Limit}

Now recall that we can obtain the result for the pure $SU(N)$ case as a $m \to \infty$ limit of the $SU(N)$ case with a single adjoint hypermultiplet matter of mass $m$ -- see (\ref{1-point to pure}). This means that we can also write (\ref{Z5d-full defect}) as
\be
Z^{{\rm pure}, \, 5d}_{{\rm inst}, \, SU(N)} ( \Lambda, \epsilon_1, 0, \vec a, T, \beta) = \langle {\bf 0} |  {\Phi}^q_{m \to \infty} (1) | {\bf 0} \rangle_{{\bf S}^2},
\ee
where ${\Phi}^q_{m \to \infty}$ is a $m \to \infty$ limit of a vertex operator of $\widehat{U_q}( {\frak {su}(N)_{\rm aff}})_{\rm crit}$, an integrable module over a quantum affine $SU(N)$-algebra at the critical level. As usual, the states $\langle {\bf 0} |$ and $| {\bf 0} \rangle$ correspond to lowest energy vertex operators ${V}^q(\infty)$ and ${V}^q(0)$ of this module inserted at $z = \infty$ and $0$, respectively. In turn, the arguments in footnote~\ref{massive limit} (which also apply to the present ``fully-ramified'', $N >1$, 5d case) mean that we can write $Z^{{\rm adj}, \, 5d}_{{\rm inst}}$ as the one-point correlation function on ${\bf T}^2$:
\be
\label{Z5d-adj-full}
\boxed{Z^{{\rm adj}, \, 5d}_{{\rm inst}, \, SU(N)} (q', \epsilon_1, 0, \vec a,  m, T, \beta)  = \langle  \,  {\Phi}^q_{m}(1)  \, \rangle_{{\bf T}^2}}
\ee
where ${\Phi}^q_{m}(z)$ is a vertex operator of $\widehat{U_q}( {\frak {su}(N)_{\rm aff}})_{\rm crit}$ (which depends on $m$), whence it defines the following map
\be
\label{Phi-m-full}
\boxed{{\Phi}^q_{m}: \, \widehat{U_q}( {\frak {su}(N)_{\rm aff}})_{\rm crit} \to \widehat{U_q}( {\frak {su}(N)_{\rm aff}})_{\rm crit}}
\ee
Clearly, in arriving at the boxed relations (\ref{Z5d-tilde}) and (\ref{tilde-gamma}), we have just derived a 5d ``fully-ramified'' AGT correspondence for $SU(N)$ with a single adjoint matter in the NS limit!

 \bigskip\noindent{\it Elliptic Macdonald Polynomials and Quantum Affine $SU(N)$-Algebra at the Critical Level}

Note that the ${\cal D}^{(l)}_{q-{\rm CM}}$'s are known to be an elliptic generalization of the Macdonald operators~\cite{Rui}. Thus, their simultaneous eigenfunctions, i.e., $Z^{{\rm adj}, \, 5d}_{{\rm inst}, \, SU(N)}$, ought to be given by an elliptic generalization of the Macdonald polynomials. In turn, (\ref{Z5d-adj-full}) and (\ref{Phi-m-full}) mean that \emph{these elliptic Macdonald polynomials  can be understood as one-point correlation functions of a 2d QFT on ${\bf T}^2$ whose underlying symmetry is generated by ${U_q}( {\frak {su}(N)_{\rm aff}})_{\rm crit}$, a quantum affine $SU(N)$-algebra at the critical level}! 

This observation is in fact consistent with various results in the mathematical literature as follows. Firstly, note that (\ref{q-CM IS}) and (\ref{Z5d-adj-full}) mean that the  ${\cal D}^{(l)}_{q-{\rm CM}}$'s correspond to central elements of  ${U_q}( {\frak {su}(N)_{\rm aff}})_{k}$ (for some level $k$) -- incidentally, this has also been pointed out in~\cite{Hasegawa} -- whence according to~\cite{Affine Mac}, they are affine Macdonald operators whose simultaneous eigenfunctions are the affine Macdonald polynomials constructed in~\cite{EK}. In turn, these affine Macdonald polynomials are given by traces of intertwiners of ${U_q}( {\frak {su}(N)_{\rm aff}})_{k}$~\cite{EK}, i.e., they are given by one-point correlation functions of vertex operators of $\widehat{U_q}( {\frak {su}(N)_{\rm aff}})_{k}$ on ${\bf T}^2$. This coincides with our aforementioned observation, after one notes the terminological correspondence between the elliptic and affine Macdonald operators and polynomials.

\newsection{An M-Theoretic Derivation of a 6d AGT Correspondence}

\newsubsection{An M-Theoretic Derivation of a 6d AGT Correspondence with Matter}

\bigskip\noindent{\it A 6d AGT Correspondence for $U(1)$ with $N_f = 2$ Fundamental Matter}

Let $n=1 = N$ in fig.~2(3), and glue together the two ends of the interval $\mathbb I_t \subset \Sigma_t = {\bf S}^1 \times {\mathbb I}_t$, where $\beta$ therein is the radius of ${\bf S}^1$. Now, $\Sigma_t$ is a two-torus ${\bf S}^1 \times {\bf S}^1_t$, and the planes at the ends labeled by $m_1$ and $m_4$ would be coincident, intersecting $\Sigma_t$ along a one-cycle in the ${\bf S}^1$-direction. On the 2d side, ${\cal C}_{\rm eff}$ also becomes a two-torus ${\bf T}^2$ after we correspondingly glue together the punctures at $z= \infty$ and $z = 0$, whence there are left two vertex operators $V_{\vec j_2}$ and $V_{\vec j_3}$. Let $\beta \nrightarrow 0$, and let the radius of ${\bf S}^1_t$ be $R_6$. Then, one can define a 6d generalization $Z^{\rm lin, \, 6d}_{\rm inst}$  of the 4d Nekrasov instanton partition function $Z^{\rm lin}_{\rm inst}$ along the $\mathbb R^4 \vert_{\epsilon_1, \epsilon_2} \times {\bf S}^1 \times {\bf S}^1_t$ worldvolume of the M5-brane lying in the $X^9\vert_{\epsilon_i}$-directions, where from (\ref{AGT-lin}), we can write
\be
\label{AGT-lin-6d}
\boxed{Z^{\rm lin, \, 6d}_{{\rm inst}, \, U(1)}(q_1, \epsilon_1, \epsilon_2, {\vec m}, \beta, R_6) = \langle \, \tilde \Phi^{w}_{v} (z_1) \, \tilde \Phi^{v}_{u} (z_2) \, \rangle_{{\bf T}^2}}
\ee
up to some constant factor, where the $N_f = 2$ fundamental matter masses are given by $\vec m = (m_1, m_2)$, and
\be
\label{variables-U(1)-6d}
\boxed{w = e^{- \beta m_1},  \quad v = e^{-2\beta (m_1 + m_2)}, \quad u = e^{- \beta m_2}} 
\ee
From a straightforward 6d generalization of our explanations in $\S$2.2 about how a 5d state would have a projection onto a circle $C_{\beta}$ in ${\cal C}_{\rm eff}$ which has radius $\beta$, we find that the vertex operator $\tilde \Phi^{c}_{d} (z)$ -- which is a 6d analog of those in (\ref{Z5d-U(1)}) -- ought to be associated with a 6d state which has a projection onto two transverse circles $C_\beta$ and $C_{R_6}$ in ${\bf T}^2$ of radius $\beta$ and $R_6$, respectively, which intersect at the point $z$.  

In the 5d case, the vertex operator was given in (\ref{VO-U(1)}). What about the 6d case? What explicit form should $\tilde \Phi^{c}_{d} (z)$ take? To answer this, first note that because $\tilde \Phi^{c}_{d} (z)$ is supposed to be associated with a state that has a projection onto the transverse circles $C_\beta$ and $C_{R_6}$ which \emph{intersect}, the commutation relation (\ref{he}) -- which is associated with the circle $C_\beta$ -- must be modified to depend on $R_6$. Since this modified commutation relation should reduce to (\ref{he}) in the limit $R_6 \to 0$, a consistent modification would be 
\be
\boxed{[{\tilde a}_m, {\tilde a}_n] = m (1 - p^{|m|}) {1- { q}^{|m|} \over 1- { t}^{|m|}} \delta_{m+n, 0}, \quad {\tilde a}_{m > 0} | \tilde \emptyset \rangle = 0}
\label{he-6d-a}
\ee
where 
\be
\label{variables-U(1)-6d-2}
\boxed{q = e^{- i \beta \sqrt{\epsilon_1 \epsilon_2}}, \quad t = e^{- i \beta ( \epsilon_1 + \epsilon_2 + \sqrt{\epsilon_1 \epsilon_2})}, \quad p = e^{- {1 \over R_6}}} 
\ee
In (\ref{he-6d-a}), we have an $|m|^{\rm th}$ power of $p$ because it is also a parameter associated with the radius of a circle in ${\bf T}^2$, i.e., it is on equal footing with $q$ and $t$. 

Second, because the 6d state has a projection onto \emph{two} transverse circles $C_\beta$ and $C_{R_6}$, it would mean that we can write an arbitrary 6d state as
\be
\label{6d state}
| \Psi \rangle = {\tilde b}_{-\mu _1} {\tilde b}_{-\mu_2} \dots {\tilde a}_{-\lambda _1} {\tilde a}_{-\lambda_2} \dots | \tilde \emptyset \rangle,
\ee
where the ${\tilde b}_{k}$'s are operators associated with the circle $C_{R_6}$, which \emph{a priori} obey the same commutation relation as the ${\tilde a}_k$'s, i.e.,
\be
{[{\tilde b}_m, {\tilde b}_n] = m (1 - p^{|m|}) {1- { q}^{|m|} \over 1- { t}^{|m|}} \delta_{m+n, 0}, \qquad {\tilde b}_{m > 0} | \tilde \emptyset \rangle = 0},
\label{he-6d-b-initial}
\ee
and 
\be
\label{a-b}
\boxed{[{\tilde a}_m, {\tilde b}_n] = 0}
\ee
Here, the $\lambda_i$'s and $\mu_i$'s define the partitions $\lambda =(\lambda_1, \lambda_2, \dots)$ and $\mu = (\mu_1, \mu_2, \dots)$.

Notice that when $\beta \to \infty$, there will not be any Omega-deformation in the directions transverse to $C_\beta$ -- recall that Omega-deformation is effected as one traverses $C_\beta$ completely, but this cannot happen if its radius is infinitely large. What this means is that the commutation relation (\ref{he-6d-b-initial}) -- which is associated with the circle $C_{R_6}$ that is transverse to $C_\beta$ -- ought to be independent of the $\epsilon_i$-dependent parameters $q$ and $t$ when $\beta \to \infty$. When $\beta \to \infty$, $q, t \to \infty$ (see (\ref{variables-U(1)-6d-2}) and footnote~\ref{parameters}); in other words, the factor of ${1- { q}^{|m|} / 1- { t}^{|m|}}$ in (\ref{he-6d-b-initial}) would tend to $(q t^{-1})^{|m|}$ as $\beta \to \infty$. Therefore, we must divide the RHS of  (\ref{he-6d-b-initial}) by $(q t^{-1})^{|m|}$ to get
\be
{[{\tilde b}_m, {\tilde b}_n] = {m (1 - p^{|m|}) \over (q t^{-1})^{|m|}} \, {1- { q}^{|m|} \over 1- { t}^{|m|}} \delta_{m+n, 0}, \qquad {\tilde b}_{m > 0} | \tilde \emptyset \rangle = 0}.
\label{he-6d-b-initial-2}  
\ee

Also, when $R_6 \to 0$, the ${\tilde b}_k$ operators should not enter our story at all; in particular, the commutation relations ought to be physically invalid when $R_6 \to 0$. Since a physically valid commutation relation would either be equal to zero or some finite constant, as $R_6 \to 0$, i.e., as $p \to 0$, the RHS of (\ref{he-6d-b-initial-2}) should blow up. In other words, we ought to modify (\ref{he-6d-b-initial-2}) to
\be
\boxed{[{\tilde b}_m, {\tilde b}_n] = {m (1 - p^{|m|}) \over (q t^{-1} p)^{|m|}} \, {1- { q}^{|m|} \over 1- { t}^{|m|}} \delta_{m+n, 0}, \qquad {\tilde b}_{m > 0} | \tilde \emptyset \rangle = 0}
\label{he-6d-b-final}
\ee
(where again, we have an $|m|^{\rm th}$ power of $p$ because it is also a parameter associated with the radius of a circle in ${\bf T}^2$, i.e., it is on equal footing with $q$ and $t$). 

Third, from (\ref{6d state}),  it is clear that we can write 
\be
\label{6d VO}
\boxed{{\tilde \Phi}^c_d (z) = {\tilde \Phi}^{c, \, {\tilde b}_k}_d (z)  \cdot {\tilde \Phi}^{c, \, {\tilde a}_k}_d (z)}
\ee
where ${\tilde \Phi}^{c, \, {\tilde a}_k ({\tilde b}_k)}_d$ is a vertex operator which involves the ${\tilde a}_k$'s (${\tilde b}_k$'s). To obtain the explicit expression for ${\tilde \Phi}^c_d$, note that according to~\cite[Theorem 1.2]{Saito}, the Fock space $\tilde {\cal F}$ spanned by the states $| \Psi \rangle$ in (\ref{6d state}) (bearing in mind the commutation relations (\ref{he-6d-a}), (\ref{he-6d-b-final}) and (\ref{a-b})), affords a representation of an elliptic generalization of the Ding-Iohara algebra we first encountered in $\S$3.2. Since the vertex operator ${\tilde \Phi}^c_d$ defines a map $\tilde{\cal F}_d \to \tilde{\cal F}_c$ from one Fock space associated with parameter $d$ to another associated with parameter $c$, by comparing the currents of the Ding-Iohara algebra in~\cite[Fact 2.6]{Awata} with the currents of the aforementioned elliptic Ding-Iohara algebra in~\cite[Theorem 1.2]{Saito}, it is clear that ${\tilde \Phi}^{c, \, {\tilde a}_k}_d$ can be obtained from (\ref{VO-U(1)}) by replacing the factor of $(1- { q}^{\pm n})$ therein with $(1 - p^{|n|})(1- { q}^{\pm n})$, i.e., 
\be
\label{VO-U(1)-6d-a}
\boxed{ \tilde\Phi^{c, \, {\tilde a}_k}_d (z) =   {\rm exp} \left(  \sum_{n > 0} {1 \over n} {c^{-n} -  d^{-n} \over  (1 - p^{n})(1- { q}^{ - n})} \, {\tilde a}_{n} \, z^{-n} \right) {\rm exp} \left( - \sum_{n > 0} {1 \over n} {c^n - (t / q)^n d^n \over  (1 - p^{n})(1- { q}^{n})} \, {\tilde a}_{-n} \, z^n \right)}  
\ee
Likewise, ${\tilde \Phi}^{c, \, {\tilde b}_k}_d$ can be obtained from (\ref{VO-U(1)}) by replacing (i) the $a_n / n$ and $- a_{-n} / n$ terms therein with $-{\tilde b}_{-n} / n$ and ${\tilde b}_{n} / n$, respectively; (ii) the factor of $(1- { q}^{\pm n})$ therein with $(1 - p^{|n|})(1- { q}^{\pm n}) / p^{|n|}$; (iii) the factor of $(1- { q}^{n}) / (t/q)^n$ therein with $(1 - p^{|n|})(1- { q}^{n}) /(t/q)^n (q/t)^{|n|} p^{|n|}$:
\be
\label{VO-U(1)-6d-b}
\boxed{ \tilde\Phi^{c, \, {\tilde b}_k}_d (z) =   {\rm exp} \left(  - \sum_{n > 0} {1 \over n} { p^{n} \, (c^{-n} -  d^{-n})  \over  (1 - p^{n})(1- { q}^{ - n})} \, {\tilde b}_{-n} \, z^{-n} \right) {\rm exp} \left(  \sum_{n > 0} {1 \over n} { p^{n} \, (c^n -  d^n) \over  (1 - p^{n})(1- { q}^{n})} \, {\tilde b}_{n} \, z^n \right)}  
\ee
One can readily check from (\ref{VO-U(1)-6d-a}), (\ref{VO-U(1)-6d-b}) and (\ref{6d VO}) that when $R_6 \to 0$ (whence $p \to 0$),  $(\tilde\Phi^{c, \, {\tilde b}_k}_d, \tilde\Phi^{c, \, {\tilde a}_k}_d ) \to (1, \Phi^{c}_d )$ so ${\tilde \Phi}^c_d \to  \Phi^{c}_d$, where $\Phi^{c}_d$ is the 5d vertex operator in  (\ref{VO-U(1)}), as expected.

Clearly, in arriving at the boxed relations (\ref{AGT-lin-6d}), (\ref{variables-U(1)-6d}), (\ref{he-6d-a}), (\ref{variables-U(1)-6d-2}), (\ref{a-b}), (\ref{he-6d-b-final}), (\ref{6d VO}), (\ref{VO-U(1)-6d-a}) and (\ref{VO-U(1)-6d-b}), we have just derived a 6d AGT correspondence for $U(1)$ with $N_f = 2$ fundamental matter!

\bigskip\noindent{\it A 6d AGT Correspondence for $SU(N)$ with $N_f = 2N$ Fundamental Matter}

Let us now proceed to consider an $SU(N)$ theory with $N_f = 2N$ fundamental matter. Note that from the 6d generalization of fig.~3, we effectively have $N$ D6-branes and $1$ D4-brane wrapping ${\cal C}_{\rm eff} = {\bf T}^2$ -- that is, we effectively have an $N \times 1 = N$-fold cover of ${\cal C}_{\rm eff}$. This means that we ought to make the following replacements to get the corresponding result for the $SU(N)$ case. 

Firstly, the vacuum state $| \tilde \emptyset \rangle$ and its dual $\langle \tilde \emptyset |$  ought to be replaced by their $N$-tensor product:
\be 
\label{Z6d-SU(N)-matter-vacuum}
{| \tilde {\bf 0} \rangle =  | \tilde\emptyset \rangle^{\otimes N} \quad {\rm and} \quad    \langle \tilde {\bf 0} | = \langle \tilde \emptyset |^{\otimes N}. }
\ee

Secondly, the vertex operators ${\tilde \Phi}^w_v$ and $ \tilde\Phi^v_u$ in (\ref{AGT-lin-6d}) ought to also be replaced by their $N$-tensor product
\be
\boxed{{\tilde \Phi}^{\bf w}_{\bf v} (z_1) = {\tilde \Phi}^{w_N}_{v_N} \cdots {\tilde \Phi}^{w_1}_{v_1} (z_1)  \qquad {\rm and} \qquad  {\tilde \Phi}^{\bf v}_{\bf u} (z_2) =   {\tilde \Phi}^{v_1}_{u_1} \cdots {\tilde \Phi}^{v_N}_{u_N} (z_2)}
\label{Z6d-SU(N)-matter-vertex operator}
\ee
where ${\bf c} = (c_1, c_2, \dots, c_N)$. 

Thus, we can express the 6d Nekrasov instanton partition function for $SU(N)$ with $N_f = 2N$ fundamental matter as
\be
\label{Z6d-SU(N)}
\boxed{{Z^{\rm lin, \, 6d}_{{\rm inst}, \, SU(N)}({q_1}, \epsilon_1, \epsilon_2, {\vec a}, {\vec m}, \beta, R_6) =  \langle  {\tilde \Phi}^{\bf w}_{\bf v} (z_1) \tilde\Phi^{\bf v}_{\bf u} (z_2)   \rangle_{{\bf T}^2}}}
\ee
where
\be
\label{Z6d-SU(N)-matter-vertex operator-map}
{{\tilde\Phi}^{\bf c}_{\bf d}: \tilde{\cal F}_{d_1} \otimes \tilde{\cal F}_{d_2} \otimes \cdots \otimes \tilde{\cal F}_{d_N} \longrightarrow \tilde{\cal F}_{c_1} \otimes \tilde{\cal F}_{c_2} \otimes \cdots \otimes \tilde{\cal F}_{c_N}}.
\ee

Note that in the $U(1)$ case, the dependence of ${\tilde\Phi}^{ c}_{d}$ on the Coulomb modulus actually drops out where it is replaced by a linear combination of the masses $m_{1,2}$ which appears in the parameter $v$ in (\ref{variables-U(1)-6d}). In other words, in the $SU(N)$ case, ${\bf v} = (v_1, \dots, v_N)$ would be associated with the Coulomb moduli ${\vec a} = (a_1, \dots, a_N)$ (where $\sum_i a_i = 0$), while ${\bf w} = (w_1, \dots, w_N)$ and ${\bf u} = (u_1, \dots, u_N)$ would continue to be associated with the $2N$ masses ${\vec m} = (m_1, m_2, \dots, m_{2N})$. As the parameters ${\bf w}, {\bf v}, {\bf u}$ must reduce to $w, v, u$ in (\ref{variables-U(1)-6d}) when $N=1$, we conclude that 
\be
\label{Z6d-SU(N)-variables}
\boxed{w_i = e^{- \beta m_i},  \quad v_i = e^{-\beta a_i}, \quad u_i = e^{- \beta m_{N+i}}}
\ee

Since $\tilde{\cal F}_{c, d}$ is a module over the elliptic Ding-Iohara algebra,  (\ref{Z6d-SU(N)-matter-vertex operator-map}) would mean that ${\tilde\Phi}^{\bf c}_{\bf d}$ is a vertex operator of an $N$-tensor product representation of the elliptic Ding-Iohara algebra. In turn, this means that 
\be
\label{ell-W}
\boxed{{\tilde\Phi}^{\bf c}_{\bf d}:  \, \widehat {{\cal W}^{q, p}_N} \to \widehat {{\cal W}^{q, p}_N}}
\ee
where $\widehat {{\cal W}^{q,p}_N}$ is a Verma module over ${\cal W}^{q,p}_N$, an elliptic affine ${\cal W}_N$-algebra.\footnote{This is true because the elliptic Ding-Iohara algebra acts on the N-tensor space ${\cal F}_1 \otimes \dots \otimes {\cal F}_N$ as ${\cal W}^{q,p}_N$, an elliptic analog of  ${\cal W}^q_{N}$, a $q$-deformed affine ${\cal W}_N$-algebra. I would like to thank Y.~Saito for an explanation of this point.}

Clearly, in arriving at the boxed relations (\ref{Z6d-SU(N)-matter-vertex operator}), (\ref{Z6d-SU(N)}),  (\ref{Z6d-SU(N)-variables}) and (\ref{ell-W}), we have just derived a 6d AGT correspondence for $SU(N)$ with $N_f = 2N$ fundamental matter! 

\bigskip\noindent{\it An Equivariant Elliptic Genus on Instanton Moduli Space and a Two-Point Correlation Function of a 2d CFT with Elliptic Affine ${\cal W}_N$-Algebra Symmetry}

Recall from our explanation in $\S$3.3 that $Z^{{\rm pure}, \, 5d}_{{\rm inst}, \, SU(N)}$ in (\ref{Z5d-pure-SU(N)}) can be regarded as a partition function associated with a (non-dynamically) $\bf g$-gauged supersymmetric quantum mechanical instanton with target  $L{\cal U}({\cal M}_{SU(N)})$, the loop space of ${\cal U}({\cal M}_{SU(N)})$.\footnote{Here, ${\bf g} \in U(1) \times U(1) \times T$, where $T \subset SU(N)$ is a Cartan subgroup.} In the same way, $Z^{\rm lin, \, 6d}_{{\rm inst}, \, SU(N)}$ can be regarded as a partition function associated with a (non-dynamically) $\bf g$-gauged supersymmetric quantum mechanical instanton  with target  $LL{\cal U}({\cal M}_{SU(N)})$, the double loop space of ${\cal U}({\cal M}_{SU(N)})$. Consequently, according to~\cite{Ba}, $Z^{\rm lin, \, 6d}_{{\rm inst}, \, SU(N)}$ ought to be given by some version of the elliptic genus of ${\cal U}({\cal M}_{SU(N)})$, whence we can also write (\ref{Z6d-SU(N)}) as (c.f.~\cite{Hollowood})
\be
 \boxed{\sum_{k = 0}^\infty q_1^k \, \chi_{{\bf g}(\epsilon_i, \vec a)}({\cal U}({\cal M}_{SU(N), k})) (\vec m, \tau) = \langle  {\tilde \Phi}^{\bf w}_{\bf v} (z_1) \tilde\Phi^{\bf v}_{\bf u} (z_2)   \rangle_{{\bf T}^2}}
\ee
where  $\chi_{\bf g}({\cal M})(\vec m, \tau)$ is the $\bf g$-equivariant elliptic genus of ${\cal M}$ which depends on the $2N$ fundamental matter masses $\vec m$ and the complex structure $\tau$ of the two-torus $\Sigma_t $ whose independent one-cycles have radii $\beta$ and $R_6$.

\newsection{A 6d ``Fully-Ramified'' AGT Correspondence and an Elliptized Integrable System}

\newsubsection{A 6d ``Fully-Ramified'' AGT Correspondence with Matter and an Elliptized Integrable System}

\bigskip\noindent{\it The ``Fully-Ramified'' 6d Nekrasov Instanton Partition Function with  $N_f = 2N$ Fundamental Matter in the NS Limit and an Elliptized Integrable System}
 
Let $Z^{\rm lin, \, 5d (6d)}_{{\rm inst}, \, SU(N)}(q_1, \epsilon_1, 0, {\vec a}, {\vec m}, \beta(,R_6), T)$ denote the ``fully-ramified'' 5d(6d) Nekrasov instanton partition function in the NS  limit $\epsilon_2 = 0$ of the $SU(N)$ theory on $\mathbb R^4\vert_{\epsilon_1, \epsilon_2 = 0} \times {\bf S}^1 (\times {\bf S}_t^1)$ with $N_f = 2N$ fundamental matter of masses $\vec m$,  Coulomb moduli $\vec a$, complexified gauge coupling $\tau'$ where $q_1 = e^{2 \pi i \tau'}$, and ${\bf S}^1$ (and ${\bf S}_t^1$) radius $\beta$ (and $R_6$), such that the $SU(N)$ gauge group reduces to its maximal torus $T$ along $\mathbb R^2\vert_{\epsilon_1} \subset \mathbb R^4\vert_{\epsilon_1, \epsilon_2 = 0}$. Let $Z^{\rm lin, \, 4d}_{{\rm inst}, \, SU(N)}(q_1, \epsilon_1, 0, {\vec a}, {\vec m}, T)$ denote the same thing in 4d over $\mathbb R^4\vert_{\epsilon_1, \epsilon_2 = 0}$. 

Note that the integrable system associated with the ``unramified'', non-Omega-deformed Nekrasov instanton partition function  $Z^{\rm lin, \, 6d}_{{\rm inst}, \, SU(N)}(q_1, 0, 0, {\vec a}, {\vec m}, \beta, R_6, SU(N))$, $Z^{\rm lin, \, 5d}_{{\rm inst}, \, SU(N)}(q_1, 0, \newline 0, {\vec a}, {\vec m}, \beta, SU(N))$ and $Z^{\rm lin, \, 4d}_{{\rm inst}, \, SU(N)}(q_1, 0, 0, {\vec a}, {\vec m}, SU(N))$, is an XYZ, XXZ and $SU(2)$ XXX spin chain, respectively~\cite{Gorsky 1}. Upon adding a full defect and turning on Omega-deformation in the NS limit, (i) according to \cite[$\S$7.2]{4d AGT}, $Z^{\rm lin, \, 4d}_{{\rm inst}, \, SU(N)}(q_1, \epsilon_1, 0, {\vec a}, {\vec m}, T) = Z^{{\rm lin}, \, 4d}_{{\rm inst}, \epsilon_1, T}$ becomes a simultaneous eigenfunction of commuting differential operators defined by the central elements of (the universal enveloping algebra of) an affine $SU(N)$-algebra at the critical level; (ii) according to (\ref{Z5d-full defect})--(\ref{Z5d-full defect-G}) and the discussion leading up to them, and our discussion in $\S$3.1 (which also applies to the 5d case with a defect), $Z^{\rm lin, \, 5d}_{{\rm inst}, \, SU(N)}(q_1, 0, 0, {\vec a}, {\vec m}, \beta, SU(N)) = Z^{\rm lin, \, 5d}_{{\rm inst}, \epsilon_1, T}$ becomes a simultaneous eigenfunction of commuting difference operators defined by the central elements of  a quantum affine $SU(N)$-algebra at the critical level; (iii) $Z^{\rm lin, \, 6d}_{{\rm inst}, \, SU(N)}(q_1, \newline 0, 0, {\vec a}, {\vec m}, \beta, R_6, SU(N)) = Z^{{\rm lin}, \, 6d}_{{\rm inst}, \epsilon_1, T}$ becomes a simultaneous eigenfunction of a set of commuting operators defined by the central elements of some algebra. These results are summarized in the diagram below. 
\tikzset{node distance=3.8cm, auto}
\begin{center}
 \begin{tikzpicture}
\node (P){$\{Z^{{\rm lin}, \, 6d}_{{\rm inst}}; \footnotesize{\text {XYZ}} \}$};
  \node (B) [right of=P] {$\{Z^{{\rm lin}, \, 5d}_{\rm inst}; \footnotesize{\text {XXZ}} \}$};
 \node (A) [right of=B] {$\{Z^{{\rm lin}, \, 4d}_{\rm inst}; \footnotesize{\text {XXX}} \}$};
  \node(P') [below of = P] {$\{Z^{{\rm lin}, \, 6d}_{{\rm inst}, \epsilon_1, T}; \,  {\rm ?} \, \}$};
  \node(B') [below of = B] [right of=P'] {$\{Z^{{\rm lin}, \, 5d}_{{\rm inst}, \epsilon_1, T}; {\cal D}^{(l)}_{{U_q}( {\frak {su}(N)_{\rm aff}})_{\rm crit}} \}$};
  \node(A') [below of = A] [right of=B'] {$\{Z^{{\rm lin}, \, 4d}_{{\rm inst}, \epsilon_1, T}; {\cal D}^{(l)}_{\frak {su}(N)_{\rm aff, crit}} \}$};
  \draw[transform canvas={yshift=-0.5ex},->] (P) --(P') node[midway] {full defect, NS};
   \draw[transform canvas={yshift=0.5ex},->] (B) --(B') node[midway] {full defect, NS};
    \draw[transform canvas={yshift=0.5ex},->] (A) --(A') node[midway] {full defect, NS};
  \draw[transform canvas={yshift=0.5ex},->] (P) --(B) node[midway] {\footnotesize{$R_6 \to 0$}};
\draw[transform canvas={yshift=-0.5ex},->](B) -- (P) node[midway] {\footnotesize{$R_6 \nrightarrow 0$}}; 
 \draw[transform canvas={yshift=0.5ex},->] (P') --(B') node[midway] {\footnotesize{$R_6 \to 0$}};
\draw[transform canvas={yshift=-0.5ex},->](B') -- (P') node[midway] {\footnotesize{$R_6 \nrightarrow 0$}}; 
\draw[transform canvas={yshift=0.5ex},->] (B) --(A) node[midway] {\footnotesize{$ {\beta} \to 0$}};
\draw[transform canvas={yshift=-0.5ex},->](A) -- (B) node[midway] {\footnotesize{${\beta} \nrightarrow 0$}};   
\draw[transform canvas={yshift=0.5ex},->] (B') --(A') node[midway] {\footnotesize{$ {\beta} \to 0$}};
\draw[transform canvas={yshift=-0.5ex},->](A') -- (B') node[midway] {\footnotesize{${\beta} \nrightarrow 0$}};   
\end{tikzpicture}
  \end{center}

We shall now argue that $Z^{{\rm lin}, \, 6d}_{{\rm inst}, \epsilon_1, T}$ is a simultaneous eigenfunction of a set of commuting operators defined by the central elements of  an elliptic affine algebra. To this end, first note that the XXX and XXZ integrable systems have underlying dynamical symmetries generated~\cite{Gorsky 1} by $Y(\frak g)$ and $U_q({\frak g}_{\rm aff})$,\footnote{Here, the Lie algebra ${\frak g} = {\frak {su}}(2)$, and ${\frak g}_{\rm aff}$ is its affine generalization.} which are a Yangian and quantum affine algebra, respectively,  such that $Y (\frak g)$ acts~\cite{Uglov} on the irreducible integrable representations of ${\frak g}_{\rm aff}$. In other words, we can associate ${\frak g}_{\rm aff}$ and $U_q({\frak g}_{\rm aff})$ with the symmetries of the XXX and XXZ integrable systems. This is reflected in the fact that  $Z^{{\rm lin}, \, 4d}_{{\rm inst}, \epsilon_1, T}$ and $Z^{{\rm lin}, \, 5d}_{{\rm inst}, \epsilon_1, T}$ are simultaneous eigenfunctions of commuting operators ${\cal D}^{(l)}_{\frak {su}(N)_{\rm aff, crit}}$ and ${\cal D}^{(l)}_{{U_q}( {\frak {su}(N)_{\rm aff}})_{\rm crit}}$ which are defined purely by an affine and quantum affine Lie algebra, respectively, as shown in the diagram above. 

Second, note that the XYZ integrable system has an underlying dynamical symmetry generated~\cite{Nuno} by $E_{\tau, \eta}(\frak g)$,\footnote{Here, the Lie algebra is again ${\frak g} = {\frak {su}}(2)$.} which is an elliptic quantum group first defined by Felder in~\cite{Felder}. Further note that the representations of $E_{\tau, \eta}(\frak g)$ are isomorphic to the representations of  $U_{p, q}({\frak g}_{\rm aff})$~\cite{Konno-work}, an elliptic affine $G$-algebra first defined by Konno in~\cite{Konno} that is a two-parameter generalization of ${U_q}( {\frak g}_{\rm aff})$, i.e., $U_{0, q}({\frak g}_{\rm aff}) = {U_q}( {\frak g}_{\rm aff})$. In sum, this means that we can associate $U_{p, q}({\frak g}_{\rm aff})$ with the symmetry of the XYZ integrable system. 

Third, let us identify $p$ with $R_6$ and $q$ with $e^\beta$. Then, as $R_6 \to 0$, we have  $U_{p, q}({\frak g}_{\rm aff}) \to {U_q}( {\frak g}_{\rm aff})$, and when $\beta \to 0$, we have ${U_q}( {\frak g}_{\rm aff}) \to U({\frak g}_{\rm aff})$, the universal enveloping algebra of ${\frak g}_{\rm aff}$. These statements hold vice-versa, so that with regard to the central elements of these algebras at the critical level, we have 

 \tikzset{node distance=3.8cm, auto}
 \begin{center}
 \begin{tikzpicture}
  \node (P) {${\cal D}^{(l)}_{U_{p, q}({\frak g}_{\rm aff})_{\rm crit}}$};
  \node (B) [right of=P] {${\cal D}^{(l)}_{{U_q}({\frak g}_{\rm aff})_{\rm crit}}$};
  \node(A) [right of=B]{${\cal D}^{(l)}_{{\frak g}_{\rm aff, crit}}$};
   \draw[transform canvas={yshift=0.5ex},->] (P) --(B) node[midway] {$R_6 \to 0$};
\draw[transform canvas={yshift=-0.5ex},->](B) -- (P) node[midway] {$R_6 \nrightarrow 0$}; 
\draw[transform canvas={yshift=0.5ex},->] (B) --(A) node[midway] {$\beta \to 0$};
\draw[transform canvas={yshift=-0.5ex},->](A) -- (B) node[midway] {$\beta \nrightarrow 0$}; 
 \end{tikzpicture}
 \end{center}
where the ${\cal D}^{(l)}_{U_{p, q}({\frak g}_{\rm aff})_{\rm crit}}$'s are a set of commuting operators defined by the central elements of  $U_{p, q}({\frak g}_{\rm aff})_{\rm crit}$, just as the ${\cal D}^{(l)}_{{U_q}({\frak g}_{\rm aff})_{\rm crit}}$'s and ${\cal D}^{(l)}_{{\frak g}_{\rm aff, crit}}$'s are sets of commuting operators defined by the central elements of  ${U_q}({\frak g}_{\rm aff})_{\rm crit}$ and $U({\frak g}_{\rm aff})_ {\rm crit}$, respectively.

Altogether, the preceding three paragraphs and diagram mean that 
\be
\label{elliptic IS}
\boxed{{\cal D}^{(l)}_{U_{p, q}(\frak {su}(N)_{\rm aff})_{\rm crit}} \cdot Z^{\rm lin, \, 6d}_{{\rm inst}, \, SU(N)}(q_1, \epsilon_1, 0, {\vec a}, {\vec m}, \beta, R_6, T) = {\cal E}^{(l)}_{U_{p, q}} Z^{\rm lin, \, 6d}_{{\rm inst}, \, SU(N)}(q_1, \epsilon_1, 0, {\vec a}, {\vec m}, \beta, R_6, T)}
\ee
where the ${\cal E}^{(l)}_{U_{p, q}}$'s are complex eigenvalues; $l = 1, 2, \dots, N-1$; and the $N-1$ commuting operators ${\cal D}^{(1)}_{U_{p, q}(\frak {su}(N)_{\rm aff})_{\rm crit}},  \dots, {\cal D}^{(N-1)}_{U_{p, q}(\frak {su}(N)_{\rm aff})_{\rm crit}}$ can be identified with the generators of a classical two-parameter-deformed affine ${\cal W}_N$-algebra~\cite{Konno-private}. As claimed, $Z^{{\rm lin}, \, 6d}_{{\rm inst}, \epsilon_1, T}$ is a simultaneous eigenfunction of a set of commuting operators defined by the central elements of  ${U_{p, q}(\frak {su}(N)_{\rm aff})_{\rm crit}}$, an elliptic affine $SU(N)$-algebra at the critical level, whence (\ref{elliptic IS}) defines an elliptized (i.e.~two-parameter deformed) integrable system.

\bigskip\noindent{\it A 6d ``Fully-Ramified'' AGT Correspondence for $SU(N)$ with $N_f = 2N$ Fundamental Matter in the NS Limit}

According to $\S$3.1 and its obvious generalization to the ``fully-ramified'' 6d case, (\ref{elliptic IS}) and (\ref{AGT-lin-6d}) would mean that
\be
\label{Z6d-full}
\boxed{Z^{\rm lin, \, 6d}_{{\rm inst}, \, SU(N)}(q_1, \epsilon_1, 0, {\vec a}, {\vec m}, \beta, R_6, T) =  \langle   {\tilde \Phi}^{{\vec m}_1}_{\vec a} (z_1) {\tilde \Phi}^{\vec a}_{{\vec m}_2} (z_2)  \rangle_{{\bf T}^2}}
\ee
where $\vec m_1 = (m_1, \dots, m_N)$ and $\vec m_2 = (m_{N+1}, \dots, m_{2N})$ are the masses of the $N_f = 2N$ fundamental matter, and  ${\tilde \Phi}^{{\vec m}_i}_{\vec a}$ is a $({\vec m}_i, \vec a)$-dependent vertex operator of $\widehat{U_{p, q}}( {\frak {su}(N)_{\rm aff}})_{\rm crit}$, an integrable module over  ${U_{p, q}(\frak {su}(N)_{\rm aff})_{\rm crit}}$, i.e., 
\be
\label{Phi-m-full-6d}
\boxed{{\tilde \Phi}^{{\vec m}_i}_{\vec a}: \, \widehat{U_{p, q}}( {\frak {su}(N)_{\rm aff}})_{\rm crit} \to \widehat{U_{p, q}}( {\frak {su}(N)_{\rm aff}})_{\rm crit}}
\ee

Clearly, in arriving at the boxed relations (\ref{Z6d-full}) and (\ref{Phi-m-full-6d}), we have just derived a 6d ``fully-ramified'' AGT correspondence for $SU(N)$ with $N_f = 2N$ fundamental matter!

\end{document}